\documentclass[letterpaper,twocolumn,aps,prb,groupedaddress,superscriptaddress,showpacs,amsfonts]{revtex4}

\usepackage{color}

\usepackage{epsfig}
\usepackage{graphicx}
\usepackage{dcolumn}
\usepackage{bm}

\DeclareMathAlphabet{\mathpzc}{OT1}{pzc}{m}{it}


\newcommand{\beq}{\begin{equation}}
\newcommand{\eeq}{\end{equation}}
\newcommand{\bea}{\begin{eqnarray}}
\newcommand{\eea}{\end{eqnarray}}

\long\def\comment#1{}

\begin{document}

\title{Solving the Parquet Equations for the Hubbard Model beyond Weak Coupling}

\author{Ka-Ming Tam}
\affiliation{Department of Physics and Astronomy, Louisiana State University, Baton Rouge, LA 70803}

\author{H. Fotso}
\affiliation{Department of Physics and Astronomy, Louisiana State University, Baton Rouge, LA 70803}

\author{S.-X. Yang}
\affiliation{Department of Physics and Astronomy, Louisiana State University, Baton Rouge, LA 70803}

\author{Tae-Woo Lee}
\affiliation{Center for Computation and Technology, Louisiana State University, Baton Rouge, LA 70803}

\author{J. Moreno}
\affiliation{Department of Physics and Astronomy, Louisiana State University, Baton Rouge, LA 70803}
\affiliation{Center for Computation and Technology, Louisiana State University, Baton Rouge, LA 70803}

\author{J. Ramanujam}
\affiliation{Department of Electrical and Computer Engineering, Louisiana State University, Baton Rouge, LA 70803}
\affiliation{Center for Computation and Technology, Louisiana State University, Baton Rouge, LA 70803}

\author{M. Jarrell}
\affiliation{Department of Physics and Astronomy, Louisiana State University, Baton Rouge, LA 70803}
\affiliation{Center for Computation and Technology, Louisiana State University, Baton Rouge, LA 70803}

\date{\today}

\date{\today}

\begin{abstract}

We find that imposing the crossing symmetry in the iteration process considerably extends the range of convergence
for solutions of the parquet equations for the Hubbard model. When the crossing symmetry is not imposed, the
convergence of both simple iteration and more complicated continuous loading (homotopy) methods are limited to high 
temperatures and weak interactions.  We modify the algorithm to impose the crossing symmetry without increasing the 
computational complexity.  We also imposed time reversal and a subset of the point group symmetries, but they did not
further improve the convergence.  We elaborate the details of the latency hiding scheme which can significantly improve 
the performance in the computational implementation.  With these modifications, stable solutions for the parquet 
equations can be obtained by iteration more quickly even for values of the interaction that are a significant fraction 
of the bandwidth and for temperatures that are much smaller than the bandwidth.  This may represent a crucial step 
towards the solution of two-particle field theories for correlated electron models.

\end{abstract}


\maketitle

\section{Introduction}

A natural step to extend most of the existing many-body
single-particle self-consistent methods is to include the full
momentum and energy dependence of the vertex corrections.
Historically, the self-consistent approach for vertex corrections was
first considered by Landau, Abrikosov and Khalatnikov in the context
of the high energy behavior of quantum
electrodynamics.\cite{Landau-Abrikosov-Khalatnikov-1954} The original
goal was to develop a non-perturbative method which encodes the
information in terms of a system of closed integral equations. The
parquet equations, in principle, provide a framework for
self-consistent determination of the self-energy and the vertex
corrections. They were proposed for both boson-boson scattering and
fermion-fermion scattering during the
1950's.\cite{Pomeranchuk-Sudakov-Ter-Martirosyan-1956,Ter-Martirosyan-1958}
Methods similar to the parquet equations were first introduced in the
context of many-body theory by de Dominics and
Martin.\cite{Dominicis-Martin-1964a,Dominicis-Martin-1964b} One of the early
practical applications was on the x-ray absorption and emission problem by Roulet, Gavoret, and
Nozi\`eres.\cite{Roulet-Gavoret-Nozieres-1969} Since then, various
problems have been studied by the parquet summation approach, most
notably, the Fermi liquid in a strong magnetic
field,\cite{Brazovskii-1972a,Brazovskii-1972b,Yakovenko-1993} the
disordered electron gas in a strong transverse magnetic
field,\cite{Kleinert-Schlegel-1995} the Anderson impurity
model,\cite{Chen-Bickers-1992,Janis-Augustinsky-2007,Janis-Augustinsky-2008,Augustinsky-Janis-2011}
random potential
problems,\cite{Janis-Kolorenc-2005,Janis-2001,Janis-2009} the Hubbard
model,\cite{Hess-Deisz-Serene-1996,Bickers-1998,Bickers-White-1991,Luo-Bickers-1993,Kusunose-2010,Janis-1999}
Helium-4,\cite{Jackson-etal-1985,Jackson-Smith-1987}
Helium-3,\cite{Pfitzner-Wolfle-1987} local moment
formation,\cite{Weiner-1970,Weiner-1971} the vortex liquid
model,\cite{Yeo-Moore-1996a,Yeo-Moore-1996b,Yeo-Moore-2001,Yeo-Park-Yi-2006}
the matrix models,\cite{Arefeva-Zubarev-1996, Shishanin-Ziyatdinov-2003}
and nuclear structure calculations.\cite{Bergli-Jensen-2010} While
these applications of parquet formulation provide a lot of important
insights, most of the calculations are based on various approximated
forms of the parquet equations.

It is obvious that going from a one particle to a two-particle
self-consistent calculation represents a significant increase in the
computational effort, as each two-particle vertex contains three
independent momentum and frequency indices.  From the point of view of
practical calculation, the number of elements for each index is around
a few thousands. Therefore, the number of elements for the vertices are 
around tens of millions to a few billions. Moreover, all the information
is encoded in integral equations with complicated structure, in which
simplification does not seem to be immediately possible. Indeed, in
the past, the most successful application using the full parquet
equations was largely limited to the single Anderson impurity model.
\cite{Chen-Bickers-1992} With recent advances in computational
infrastructure where peta-scale performance has become
available, calculations for lattice models, such as the
Hubbard model are now feasible. For example, the solution of the
parquet equations for a $4 \times 4$ Hubbard cluster with on-site
coupling $U=2t$ and temperature $T=0.3t$ was recently
obtained.\cite{Yang-etal-2009}

However, limitations on computer performance and storage are
apparently not the sole obstacles for obtaining the solution of the
parquet equations. Another major barrier is the stability of the
solvers. The simple iteration method, which is widely adapted for the
dynamical mean field method, often fails to 
provide a stable solution for the parquet equations. In most cases, a
damping scheme has to be employed. Even with the damping scheme, when
the temperature is low or the coupling is large, finding a stable
solution still seems to be rather difficult. \cite{Yang-etal-2009} 

Given the large number of variables and the complexity of the parquet
equations, instabilities in their solution may not be unexpected. Methods 
based on the local gradient are not likely to be suitable as the Hessian
cannot be readily calculated. Most of the non-linear solvers only have
local convergence properties.  This may not pose a problem if we have a
reasonable guess which is close enough to the true
solution. Unfortunately, it is not easy to obtain a good initial guess
for the parquet equations. Methods that in principle allow ``global"
convergence, for example, the homotopy method or continuous loading method, 
have been proposed as possible ways to improve the
calculations.\cite{Bickers-homotopy} While these tend to improve
convergence, many steps are required for the solution to move along
the homotopy path. Moreover, practical experience seems to suggest
that convergence may still not be achieved when the temperature is low
or the coupling is strong. It is clear that a better solver is
definitely required for the practical application of the parquet
method within the context of the strongly correlated systems.

One of the most prominent differences between the parquet formulation
and most of the other approximation schemes such as RPA,\cite{Bohm-Pines-1953,Pines-1953}
self-consistent spin fluctuations approach,\cite{Moriya-1985} and fluctuation exchange
approach\cite{Bickers-Scalapino-1989} is that the so-called crossing symmetry is obeyed by construction
of the parquet equations. The crossing symmetry\cite{Weinberg-book1} implies that a 
vertex in one channel can also produce the vertex in all other
channels by pulling or crossing the vertex legs and multiplying 
by appropriate constants. It also implies the Pauli
exclusion principle is automatically satisfied. However, in the
course of the iteration process, as long as the exact solution of
the parquet equations is not obtained, the crossing symmetry is
violated.  The main point in the present paper is to highlight that
the crossing symmetry is crucial for obtaining a stable solution. We
devise a modified iteration scheme which can obtain a stable solution
for the parquet equations at lower temperature and stronger coupling
than that from the previous schemes.\cite{Yang-etal-2009} This is achieved primarily by
restoring the crossing symmetry at each step of the iteration.

It is important to notice that because of the large number of vertex
functions, for production runs, massively parallel machines are
absolutely necessary. Since the vertex functions in different channels
are mixed in the parquet formulation, an efficient scheme to
transform the vertex functions storage in different nodes is critical
to improving the overall efficiency of the calculations. Some of the
computational details have been explained in the previous
publication.\cite{Yang-etal-2009} We have further improved the scheme
which allows us to hide the communication latency across different
nodes behind the local calculations within the nodes, which effectively
further speeds up the calculations.

The model used for testing the computational
scheme for solving the parquet equations is the standard
Hubbard model at half-filling. The Hamiltonian is
\begin{eqnarray}
H &=& -t\sum_{<i,j>}(c_{i,\sigma}^{\dagger}c_{j,\sigma} + H.c.) +
U\sum_{i}n_{i,\uparrow}n_{i,\downarrow},
\end{eqnarray}
where $U$ is the on-site repulsion and $t=1$ is the hopping matrix which
establishes a unit of energy.  

The paper is organized as follows. In Section II, we reproduce the parquet 
equation which also serves to fix the notation. In Section III, we describe 
the iteration scheme for solving the parquet equations. In Section IV, we discuss 
the violation of crossing symmetry. In Section V, we present a modified iteration 
scheme which explicitly restores the crossing symmetry. We also discuss the 
limitation of the modified scheme and the possible directions for further
developments. In Section VI, we present the leading eigenvalues of the antiferromagnetic channel
as a function of the temperature and coupling strength and find that the parameter region of
stable solutions is greatly increased when the crossing symmetry is enforced.
A brief summary is contained in Section VII.
In the Appendix, we present a latency hiding scheme
which allows substantial increase of the efficiency for solving the
parquet equations.

\section{Parquet equations}

In order for the current paper to be reasonably self-contained, we
provide in this section a brief description of the parquet equations,
largely following Ref. \onlinecite{Yang-etal-2009}.  The purpose is
to highlight the structure of the equations and to define the
notations that will be used in the subsequent sections. We do not
intend to provide a derivation of the formulation. For this we
refer the readers to the literature where ample discussions can be
found.\cite{Bickers-Scalapino-1992,Kleinert-2009,Yang-etal-2009,Bickers-1998,Bickers-White-1991,Kusunose-2010}

Standard perturbative expansions attempt to describe all the
scattering processes, at the lowest orders, as single- or two-particle Feynman diagrams. In
the single-particle formulation the self-energy describes the many-body
processes that renormalize the motion of a particle in the interacting
background of all the other particles. In the two-particle context,
one is able to probe the interactions between particles using the so-called
vertex functions, which are matrices describing the two particle
scattering processes. For example, the reducible (full) two-particle vertex
$F^{ph}(12;34)$ describes the amplitude of a particle-hole pair
scattered from its initial state $\left|3,4\right>$ into the final
state $\left|1,2\right>$. Here, $i = 1,2,3,4$ represents a set of
indices which combines the momentum $\mathbf{k}_i$, the Matsubara
frequency $i\omega_{n_i}$ and, if needed, the spin $\sigma_i$ and band
index $m_i$. Since the total momentum and energy of the vertex are conserved,
it is convenient to adapt the notation $F^{ph}(2-3)_{1,3}$ for the numerical 
implementation of the single band Hubbard model. \cite{Yang-etal-2009}

In general, depending on how particles or holes are involved in the
scattering processes, one can define three different two-particle
scattering channels. These are the particle-hole (p-h) horizontal
channel, the p-h vertical channel and the particle-particle (p-p)
channel. 

One can further
discriminate the vertices according to 
their topology. Starting from the 
reducible vertex $F$ introduced above, we may define the irreducible
vertex $\Gamma $ corresponding to the subclass of diagrams in $F$ that
cannot be separated into two parts by cutting
two horizontal Green function lines. Similarly, the fully
irreducible vertex $\Lambda$ corresponds to the subclass of diagrams
in $\Gamma $ that cannot be split into two parts by cutting
two Green function lines in any channel.

Furthermore, since we are mostly interested in models that preserve the
$SU(2)$ spin rotation symmetry, and since this is an exact symmetry for our
two-dimensional calculations at non-zero temperature, it is convenient
to preserve this symmetry. This is accomplished by decomposing the
vertices in the so-called spin-diagonalized
representation.\cite{Bickers-1998,Bickers-White-1991} In this
representation, the spin degrees of freedom decompose the
particle-hole channel into the density and the magnetic channels, and
the particle-particle channel into the spin singlet and the spin
triplet channels which we denote as $d$-channel, $m$-channel,
$s$-channel, and $t$-channel respectively.  They are defined as
follows,

\begin{eqnarray}
\Gamma_{d} &=& \Gamma^{PH}_{\uparrow \uparrow ; \uparrow \uparrow} + \Gamma^{PH}_{\uparrow \uparrow ; \downarrow \downarrow}, \\
\Gamma_{m} &=& \Gamma^{PH}_{\uparrow \uparrow ; \uparrow \uparrow} - \Gamma^{PH}_{\uparrow \uparrow ; \downarrow \downarrow}, \\
\Gamma_{s} &=& \Gamma^{PP}_{\uparrow \downarrow ; \uparrow \downarrow} - \Gamma^{PP}_{\uparrow \downarrow ; \downarrow \uparrow}, \\
\Gamma_{t} &=& \Gamma^{PP}_{\uparrow \downarrow ; \uparrow \downarrow} + \Gamma^{PP}_{\uparrow \downarrow ; \downarrow \uparrow},
\end{eqnarray}
and similarly for $F$ and $\Lambda$.

We reproduce the full set of equations for the parquet formulation in
the spin diagonalized representation in the
following.\cite{Yang-etal-2009,Kusunose-2010,Bickers-1998} The
Schwinger-Dyson equation is
\begin{widetext}
\begin{eqnarray}
\Sigma(P) &=& -\frac{UT^2}{4N} \sum_{P^\prime,Q} \{G(P^\prime) G(P^\prime+Q)
G(P-Q) (F_d(Q)_{P-Q,P^\prime}-F_m(Q)_{P-Q,P^\prime}){\nonumber} \\
&& \;\;\;\;\;\;\;\;\;\;\;\;\;\; + G(-P^\prime) G(P^\prime+Q) G(-P+Q)
(F_s(Q)_{P-Q,P^\prime}+F_t(Q)_{P-Q,P^\prime})\},
\end{eqnarray}
\end{widetext}
where $G$ is the single-particle Green function, which itself can be
calculated from the self-energy using the Dyson equation,
\begin{eqnarray}
G^{-1}(P) &=& G_0^{-1}(P)\; - \; \Sigma(P),
\end{eqnarray}
where $G_{0}$ is the bare Green function. Here, the indices $P$, $P^\prime$ and $Q$ combine momentum ${\bf k}$
and Matsubara frequency $i\omega_n$, i.e.\ $P=({\bf k},i\omega_n)$.

The reducible and the irreducible vertices in a given channel are
related by the Bethe-Salpeter equation,
\begin{equation}
F_{r}(Q)_{P,P^\prime} = \Gamma_{r}(Q)_{P,P^\prime} +
\Phi_{r}(Q)_{P,P^\prime}, 
\label{SD_BSPH_EQ}
\end{equation}
\begin{equation}
F_{r^\prime}(Q)_{P,P^\prime} = \Gamma_{r^\prime}(Q)_{P,P^\prime} +
\Psi_{r^\prime}(Q)_{P,P^\prime},
\label{SD_BSPP_EQ}
\end{equation}
where $r=d\;\mbox{or}\;m$ for the density and magnetic channels
and $r^\prime=s\;\mbox{or}\;t$ for the spin singlet and
spin triplet channels. The vertex ladders are
defined as
\begin{eqnarray}
\Phi_{r}(Q)_{P,P^\prime} \equiv 
\;\;\;\;\;\;\;\;\;\;\;\;\;\;\;\;\;\;\;\;\;\;\;\;\;\;\;\;\;\;\;\;\;\;\;\;\;\;\;\;\;\;\;\;\;
\label{eq:Phi_ladder} \\ \nonumber
\sum_{P^{\prime\prime}}F_{r}(Q)_{P,P^{\prime\prime}}\chi_0^{ph}(Q)_{P^{%
\prime\prime}} \Gamma_{r}(Q)_{P^{\prime\prime},P^\prime}, \\
\Psi_{r^\prime}(Q)_{P,P^\prime} \equiv 
\;\;\;\;\;\;\;\;\;\;\;\;\;\;\;\;\;\;\;\;\;\;\;\;\;\;\;\;\;\;\;\;\;\;\;\;\;\;\;\;\;\;\;\;\;
\label{eq:Psi_ladder}  \\ \nonumber
\sum_{P^{\prime\prime}}F_{r^\prime}(Q)_{P,P^{\prime\prime}}\chi_0^{pp}(Q)_{P^{%
\prime\prime}} \Gamma_{r^\prime}(Q)_{P^{\prime\prime},P^\prime},
\end{eqnarray}
where $\chi_0$ is the product of two single-particle Green
functions.  

The parquet equations in the spin diagonalized representation are 
\begin{widetext}
\begin{eqnarray}
\label{Gamma_d}
\Gamma_d(Q)_{P{P^\prime}} &=& \Lambda_d(Q)_{P{P^\prime}} - {\frac{1 }{2}}%
\Phi_d({P^\prime}-P)_{P,P+Q} - {\frac{3 }{2}}\Phi_m({P^\prime}-P)_{P,P+Q} 
\\
&& \;\;\;\;\;\;\;\;\;\;\;\;\;\;\; + \; {\frac{1 }{2}}\Psi_s(P+{P^\prime}%
+Q)_{-P-Q,-P} + {\frac{3 }{2}}\Psi_t(P+{P^\prime}+Q)_{-P-Q,-P}, \nonumber
\end{eqnarray}
\begin{eqnarray}
\label{Gamma_m}
\Gamma_m(Q)_{P{P^\prime}} &=& \Lambda_m(Q)_{P{P^\prime}} - {\frac{1 }{2}}%
\Phi_d({P^\prime}-P)_{P,P+Q} + {\frac{1 }{2}}\Phi_m({P^\prime}-P)_{P,P+Q} 
\\
&& \;\;\;\;\;\;\;\;\;\;\;\;\;\;\;\; - \; {\frac{1 }{2}}\Psi_s(P+{P^\prime}%
+Q)_{-P-Q,-P} + {\frac{1 }{2}}\Psi_t(P+{P^\prime}+Q)_{-P-Q,-P}, \nonumber
\end{eqnarray}
\begin{eqnarray}
\label{Gamma_s}
\Gamma_s(Q)_{P{P^\prime}} &=& \Lambda_s (Q)_{P{P^\prime}} + {\frac{1 }{2}}%
\Phi_d({P^\prime}-P)_{-{P^\prime},P+Q} - {\frac{3 }{2}}\Phi_m({P^\prime}%
-P)_{-{P^\prime},P+Q}  \\
&& \;\;\;\;\;\;\;\;\;\;\;\;\;\;\; + \; {\frac{1 }{2}}\Phi_d(P+{P^\prime}%
+Q)_{-{P^\prime},-P} - {\frac{3 }{2}}\Phi_m(P+{P^\prime}+Q)_{-{P^\prime},-P}, \nonumber
\end{eqnarray}
\begin{eqnarray}
\label{Gamma_t}
\Gamma_t(Q)_{P{P^\prime}} &=& \Lambda_t (Q)_{P{P^\prime}} + {\frac{1 }{2}}%
\Phi_d({P^\prime}-P)_{-{P^\prime},P+Q} + {\frac{1 }{2}}\Phi_m({P^\prime}%
-P)_{-{P^\prime},P+Q}  \\
&& \;\;\;\;\;\;\;\;\;\;\;\;\;\;\; - \; {\frac{1 }{2}}\Phi_d(P+{P^\prime}%
+Q)_{-{P^\prime},-P} - {\frac{1 }{2}}\Phi_m(P+{P^\prime}+Q)_{-{P^\prime},-P}. \nonumber
\end{eqnarray}
\end{widetext}

It is important to note at this point that if we substitute the
irreducible vertices $\Gamma$
(Eqs.~\ref{Gamma_d},\ref{Gamma_m},\ref{Gamma_s}, and \ref{Gamma_t}) into the Bethe-Salpeter
equation (Eqs.~\ref{SD_BSPH_EQ} and \ref{SD_BSPP_EQ}) the crossing symmetry in
the full vertex $F$ is automatically satisfied regardless of the
numerical values of the vertex ladders $\Phi$ and $\Psi$, assuming the
fully irreducible vertices, $\Lambda$, are crossing symmetric. We write all the
full vertices explicitly in the following using only the vertex ladders, $\Phi$, $\Psi$,
and the fully irreducible vertices, $\Lambda$.

\begin{widetext}

\begin{eqnarray}
\label{F_d}
F_{d}(Q)_{P,P^\prime} = \Lambda_d(Q)_{P{P^\prime}} - {\frac{1 }{2}} \Phi_d({P^\prime}-P)_{P,P+Q} - {\frac{3 }{2}}\Phi_m({P^\prime}-P)_{P,P+Q} \\ 
+ {\frac{1 }{2}}\Psi_s(P+{P^\prime}+Q)_{-P-Q,-P} + {\frac{3 }{2}}\Psi_t(P+{P^\prime}+Q)_{-P-Q,-P} + \Phi_{d}(Q)_{P,P^\prime}; \nonumber
\end{eqnarray}

\begin{eqnarray}
\label{F_m}
F_{m}(Q)_{P,P^\prime} &=& \Lambda_m(Q)_{P{P^\prime}} - {\frac{1 }{2}}\Phi_d({P^\prime}-P)_{P,P+Q} + {\frac{1 }{2}}\Phi_m({P^\prime}-P)_{P,P+Q} \\ 
&-&  {\frac{1 }{2}}\Psi_s(P+{P^\prime}+Q)_{-P-Q,-P} + {\frac{1 }{2}}\Psi_t(P+{P^\prime}+Q)_{-P-Q,-P} + \Phi_{m}(Q)_{P,P^\prime}; \nonumber
\end{eqnarray}

\begin{eqnarray}
\label{F_s}
F_{s}(Q)_{P,P^\prime} &=& \Lambda_s (Q)_{P{P^\prime}} + {\frac{1 }{2}} \Phi_d({P^\prime}-P)_{-{P^\prime},P+Q} - {\frac{3 }{2}}\Phi_m({P^\prime}-P)_{-{P^\prime},P+Q}  \\ 
&+&  {\frac{1 }{2}}\Phi_d(P+{P^\prime}+Q)_{-{P^\prime},-P} - {\frac{3 }{2}}\Phi_m(P+{P^\prime}+Q)_{-{P^\prime},-P} + \Psi_{s}(Q)_{P,P^\prime}; \nonumber
\end{eqnarray}

\begin{eqnarray}
\label{F_t}
F_{t}(Q)_{P,P^\prime} &=& \Lambda_t (Q)_{P{P^\prime}} + {\frac{1 }{2}} \Phi_d({P^\prime}-P)_{-{P^\prime},P+Q} + {\frac{1 }{2}}\Phi_m({P^\prime}-P)_{-{P^\prime},P+Q} \\  
&-& {\frac{1 }{2}}\Phi_d(P+{P^\prime}+Q)_{-{P^\prime},-P} - {\frac{1 }{2}}\Phi_m(P+{P^\prime}+Q)_{-{P^\prime},-P} + \Psi_{t}(Q)_{P,P^\prime}. \nonumber
\end{eqnarray}

\end{widetext}
These relations allow us to restore the crossing symmetry for the full
vertices without heavy computational overhead.

The prominent technical problem at hand is whether or not we can solve
this set of equations efficiently without resorting to any
approximated scheme. An obvious difficulty is to handle the large
number of variables. On going from the one-particle level calculation to two-particle
level calculation, the number of variables which has to be
monitored grows as the third power of the linear dimension of the
system. If $N_{t}$ is the number of lattice sites times the number of discrete Matsubara
frequencies, i.e., $N_{t} = N_{k} \times N_{\omega}$, the largest $N_{t}$ that can be handled 
is in the range $2000-3000,$ i.e., the number of variables can be over one billion.  
One can immediately see that practical calculations for
reasonably large system sizes pose a serious problem, although not
insurmountable with modern computational facilities where a large  number of computer nodes 
are accessible. However, in addition to the large number of computations associated
with solving the parquet equations, they also require a complex communication pattern
between the different processes as we discussed in more detail in the Appendix.  
Moreover, a numerical instability is not unexpected, especially when the system 
in the proximity of a phase transition.

\section{Numerical implementation}

The parquet formulation consists of two sets of equations. The first
set, made of the parquet equations and the Bethe-Salpeter equation, determines
the full vertex $F$ and the irreducible vertex $\Gamma$ given the 
one-particle self-energy $\Sigma$ and the fully irreducible vertex
$\Lambda$ as the inputs. The second set of equations determine the 
one-particle quantities given the full vertex $F$; it includes the 
Schwinger-Dyson equation and the Dyson equation.

Since the method is iteration based, the initial guess is crucial for 
obtaining a converged solution. In principle, the initial guess can be 
approximated, for example, by second order perturbation. However, 
in practice, this is not the optimal choice, especially when the 
self-energy from the second order perturbation is small.  In this case 
the Green function will quickly destabilize the calculation.  This may 
relate to the fact that the damping from the imaginary part of the 
self-energy is quickly reduced. Since we are supposing that we know 
the fully irreducible vertex $\Lambda$ , a practical scheme is to choose the 
irreducible, $\Gamma$ and full vertices, $F$,  equal to $\Lambda$, and a large value
(a few times of the bandwidth) for the imaginary part of the self-energy.

\begin{figure}
\centerline{
\includegraphics*[height=0.30\textheight,width=0.45\textwidth, viewport=0 140 600 510, clip]{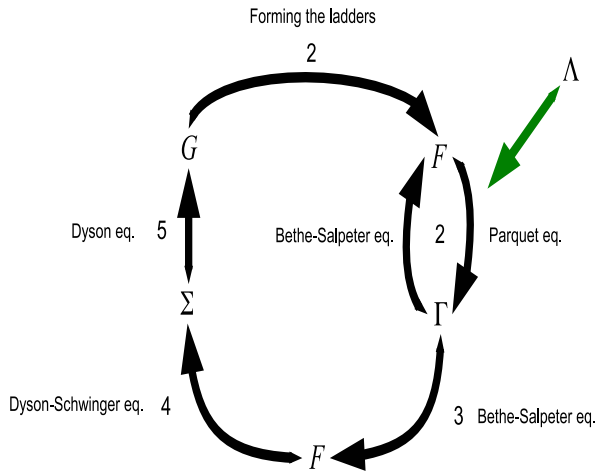}}
\caption{(Color online) Flow diagram of the algorithm for solving the parquet equations. See the text for the description of each step.
The major computational bottleneck is in the self-consistent loop of step 2. The cross channel rotations of the vertex 
ladders from the form required by the Bethe-Salpeter equations to that in the parquet equations require expensive 
communications across different nodes in the parallel implementation.}
\label{algorithm_A}
\end{figure}

Fig.~\ref{algorithm_A} is an illustration of the flow diagram of the algorithm, 
where the fully irreducible vertices are the initial input for the calculation.
The algorithm can be described as the following:

1. Set the initial conditions for the irreducible and full vertices, and the self-energy. 

2. Update the Green's functions and calculate the bare susceptibility, $\chi_0$, which is given 
by the product of two Green's functions. Solve the parquet and the Bethe-Salpeter equations for 
the irreducible vertices, $\Gamma$. Simple iteration is used until the convergence criteria are met for the
irreducible vertices.

This completes the update of the vertices. The next step is to use the
irreducible vertices obtained from the parquet equations to construct the
full vertices.

3. Solve the Bethe-Salpeter equation to obtain the full vertices, $F$, 
using the irreducible vertices from the previous step; this is
executed exactly by calling the LAPACK routines for the inverse of the
matrices. \cite{lapack}

With the full vertices obtained, we can update the self-energy.

4. Solve the Dyson-Schwinger equation to obtain the self-energy
from the full vertices. Simple iteration is used until certain
convergence criteria are met for the self-energy.

5. Solve the Dyson equation for the fully dressed Green function
from the self-energy. 

This completes the iteration loop, and the procedure is repeated from
step 2, until convergence is reached for both the
self-energy and the irreducible vertices.

In practice, step 2 which attempts to obtain the irreducible
vertices needs to be iterated for a few times to get a
reasonable convergence, even in the case where the coupling is weak
and the temperature is high. On the other hand, step 4 which
attempts to solve the self-energy from the updated full vertices is
not iterated more than one time at each loop, so as to avoid
instability (we define instability here as the failure to obtain a
converged solution from the iterative solver). Attempting to solve the self-energy at the early stage of
the iteration procedure where the full vertices are not well converged
can generally lead to instability. Although in the present paper, we
only focus on the Hubbard model,  instabilities in the iteration
process have also been observed in solving the parquet equations for 
nuclear structure calculations.\cite{Bergli-Jensen-2010}

A widely used method to avoid the instability in the iterative process
is to introduce a damping factor in the updates of the variables. The
updates are modified as follows.
\begin{eqnarray}
\Sigma &=& (\alpha)  \Sigma_{new} + (1-\alpha) \Sigma_{old}; \\
\Gamma &=& (\alpha)  \Gamma_{new} + (1-\alpha) \Gamma_{old}.
\end{eqnarray}
With this damping scheme, the solution for the half-filled Hubbard
model on a $4 \times 4 $ cluster has been successfully obtained for
$U=2t$ and temperature, $T=0.3t$.\cite{Yang-etal-2009} However, in
the strong coupling regime, obtaining a stable solution still seems to
be difficult, even though a rather heavy damping is employed.

We will demonstrate the instability problem of the simple iterative
process by monitoring the leading eigenvalues $\lambda$ defined as
\begin{equation}
\lambda_{r} \phi_{r} = \Gamma_{r}(P,P^{'}) G(P^{'})G(P^{'}+(\pi,\pi))\phi_{r}
\label{eigen_ph}
\end{equation}
for $r=d$ and $m$; similarly
\begin{equation}
\lambda_{r^\prime} \phi_{r^\prime} = (-1/2)\Gamma_{r^\prime}(P,P^{'}) G(-P^{'})G(P^{'})\phi_{r^\prime}
\label{eigen_pp}
\end{equation} 
for $r^\prime=s$ and $t$.
In principle, these leading eigenvalues signal a phase
transition by going through $1$, expressing the divergence of the
susceptibilities in the corresponding channel.

In Fig.~\ref{nosym}, we plot the leading eigenvalues of the density, 
magnetic, spin singlet, and spin triplet channels as a function of the 
number of iteration steps calculated with this simple iteration method (SI).
The calculation is done on a 2$\times$2 cluster with 32 frequencies and 
temperature $T=0.4t$; the damping parameter is $\alpha=0.1$. A converged 
solution is obtained for $U=2t$; however, for $U=4t$ and $6t$ the iterative 
solutions diverge. Changing the damping or the initial self-energy
does not help in obtaining a converged
solution for the larger values of $U$.  These are illustrative examples which
show the problem of using the simple iteration method for solving the
parquet equations. For weak coupling and not too low temperature,
converged solution can be obtained. Beyond weak coupling the iteration
becomes divergent.

\subsection{Continuous loading method}
A widely used method to alleviate the divergence in the non-linear solver is the so-called 
continuous loading or homotopy method. The basis of the continuous loading method is to 
construct an auxiliary equation with a tuning parameter $\nu$, so that its solution is trivial 
for $\nu=0$ but the solution of the original equation is recovered for $\nu=1$. Symbolically 
we can write down the set of equations of the parquet formulation as
${\bf f}_{parquet}(\Sigma,\Gamma)= 0$, where ${\bf f}_{parquet}$ is a large vector. 
We define the auxiliary function as ${\bf g}=\nu {\bf f}_{parquet}+(1-\nu){\bf f}_{0}$, where  
${\bf f}_{0}$ is a function with a trivial solution.  In our study we choose ${\bf f_{0}}$ as 
a vector containing all the elements of $\Gamma - \Gamma_{0}$ and
$\Sigma - \Sigma_{0}$, where $\Gamma_{0}$ and $\Sigma_{0}$ are the initial guesses for
the irreducible vertices and the self-energy respectively. The iteration method is used to solve the
function ${\bf g}(\nu)$, instead of the ${\bf f}_{parquet}$. One can readily see that the solution of 
${\bf g}(0)$ is trivial while the solution of the ${\bf f}_{parquet}$ is recovered when $\nu=1$. The idea is to find a
converged solution for ${\bf g}(\nu)$ with a small enough $\nu$ where a
converged solution can be readily obtained, and then gradually increase $\nu$
until it goes to $1$. Therefore, a series of $\nu$ values are needed,
which we denote as $\nu_{i}$. 

We plot the leading eigenvalues from
the continuous loading method in Fig.~\ref{continue_nosym}.  
The values of $\nu$ used are $\nu=0.0,0.5,0.8,0.9,0.95,0.99,0.993,0.996,0.997,0.998,0.999,$ $0.9999,1.0$. 
$100$ and $200$ iterations are performed for $\nu \leq
0.993$ and $\nu \geq0 .996$ respectively. 
For $U=2t$, converged solution is obtained for
$\nu=1$. However, for $U=4t$ and $6t$, the iterative solution 
diverges. The divergences appear before the homotopy parameter is
pushed to $\nu=1$. These examples illustrate the generic behavior
when solving the parquet equations by the simple iteration method beyond
weak coupling. They also illustrate that the continuous loading method
may not be sufficiently robust to solve the problem.  The damping
factor $\alpha$ used in these calculations is $0.1$ which we believe
is a fairly small value, although it may still not be sufficient. A 
rule of thumb for choosing the damping parameter is that the damping
parameter should be close to the value of the inverse of the
residual between two consecutive iterations; unfortunately, with the huge number
of variables, this choice will result in a very small step and may not
be a practical option.\cite{Bank-Rose-1981}

\section{Crossing Symmetry Violation}

\begin{figure}[t]
\includegraphics*[width=8cm]{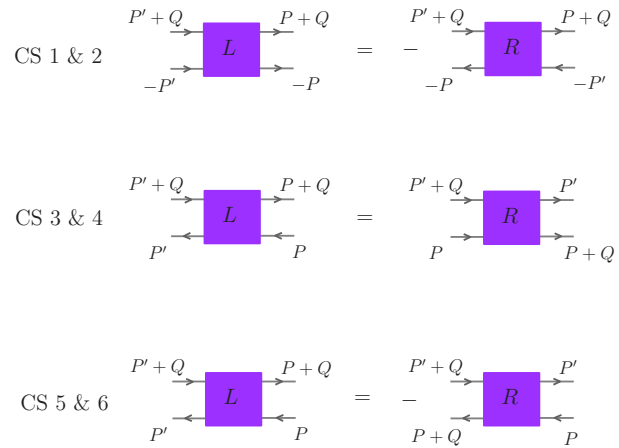}
\caption{(Color online) Diagrammatic representation of the six crossing symmetry operations. 
Note that spin indices are hidden for the purpose of clarity. For the first two operations
(CS 1 \& 2), we exchange the lower two external legs. For third and fourth operations (CS 3 \& 4), 
we exchange the lower two legs and then the right two legs.   The first four crossing symmetry 
relationships (CS 1 -- 4) relate the particle-particle vertices with the particle-hole vertices.   
The last two (CS 5 \& 6) are for the particle-hole vertices only, where we exchange the lower 
left and upper right legs. 
}
\label{crossing_symmetries}
\end{figure}

The exact solution of the parquet equations automatically satisfies the
crossing symmetry. It is one of the most important differences between
the parquet formulation and most other perturbative methods. However,
within the iteration scheme presented in the last section, the crossing symmetry is not
fulfilled unless the iteration converges to an exact solution. From
the above section, we clearly find that the iteration method, even with
the help of the continuous loading scheme, is not robust enough to obtain a
converged solution beyond weak coupling. It is desirable to quantify
the violation of crossing symmetry.  The following six equalities are
the consequence of the crossing symmetry (see Fig. \ref{crossing_symmetries} 
for a diagrammatic representation of these crossing symmetry (CS) operations). 
\cite{Bickers-1998,Kusunose-2010}
\begin{eqnarray}
L_{1} \equiv F_{t}(Q)_{P,P^{'}} = 
\label{crossing_equalities1}
\;\;\;\;\;\;\;\;\;\;\;\;\;\;\;\;\;\;\;\;\;\;\;\;\;\;\;\;\;\;\;\;\;\;\;\;\;\;\;\;\;\;\;\;\;\; \\ \nonumber
[-(1/2) F_{m}-(1/2)F_{d}](P+P^{'}+Q)_{-P^{'},-P} \equiv R_{1},  \\
L_{2} \equiv F_{s}(Q)_{P,P^{'}} = 
\;\;\;\;\;\;\;\;\;\;\;\;\;\;\;\;\;\;\;\;\;\;\;\;\;\;\;\;\;\;\;\;\;\;\;\;\;\;\;\;\;\;\;\;\;\; \\ \nonumber
[-(3/2) F_{m}+(1/2)F_{d}](P+P^{'}+Q)_{-P^{'},-P} \equiv R_{2},  \\
L_{3} \equiv F_{m}(Q)_{P,P^{'}} = 
\;\;\;\;\;\;\;\;\;\;\;\;\;\;\;\;\;\;\;\;\;\;\;\;\;\;\;\;\;\;\;\;\;\;\;\;\;\;\;\;\;\;\;\;\;\; \\ \nonumber
[(1/2)F_{t}-(1/2)F_{t}](P+P^{'}+Q)_{-P-Q,-P} \equiv R_{3}, \;\\
L_{4} \equiv F_{d}(Q)_{P,P^{'}} = 
\;\;\;\;\;\;\;\;\;\;\;\;\;\;\;\;\;\;\;\;\;\;\;\;\;\;\;\;\;\;\;\;\;\;\;\;\;\;\;\;\;\;\;\;\;\; \\ \nonumber
[(3/2)F_{t}+(1/2)F_{t}](P+P^{'}+Q)_{-P-Q,-P} \equiv R_{4}, \;\\
L_{5} \equiv F_{m}(Q)_{P,P^{'}} = 
\;\;\;\;\;\;\;\;\;\;\;\;\;\;\;\;\;\;\;\;\;\;\;\;\;\;\;\;\;\;\;\;\;\;\;\;\;\;\;\;\;\;\;\;\;\; \\ \nonumber
[(1/2)F_{m}-(1/2)F_{d}](P^{'}-P)_{P,P+Q} \equiv R_{5}, \;\;\;\;\;\;\;\;\;\;\; \\
L_{6} \equiv F_{m}(Q)_{P,P^{'}} = 
\label{crossing_equalities6}
\;\;\;\;\;\;\;\;\;\;\;\;\;\;\;\;\;\;\;\;\;\;\;\;\;\;\;\;\;\;\;\;\;\;\;\;\;\;\;\;\;\;\;\;\;\; \\ \nonumber
[-(3/2)F_{m}+(1/2)F_{d}](P^{'}-P)_{P,P+Q} \equiv R_{6}. \;\;\;\;\;\;\;\; \nonumber 
\end{eqnarray}

In Fig.~\ref{nosym_crossing} we plot the violation of crossing
symmetry versus the number of iterations.  It can be seen clearly that the
crossing symmetry cannot be perfectly restored, even for the case of $U = 2t$,
the measures of violation of crossing show oscillatory decreasing
behavior, and the rate of decrease is quite slow even though the
leading eigenvalues seem to be well converged. Obviously, for the cases
of $U=4t$ and $6t$ the crossing symmetry is severely violated and at
the verge of the divergence there are sharp increases in the crossing
symmetry violation. This may suggest that if the crossing symmetry can
be restored, the divergence may be avoided beyond weak coupling. 

\begin{figure}
\centerline{
\includegraphics*[height=0.31\textheight,width=0.26\textwidth, angle=270, clip]{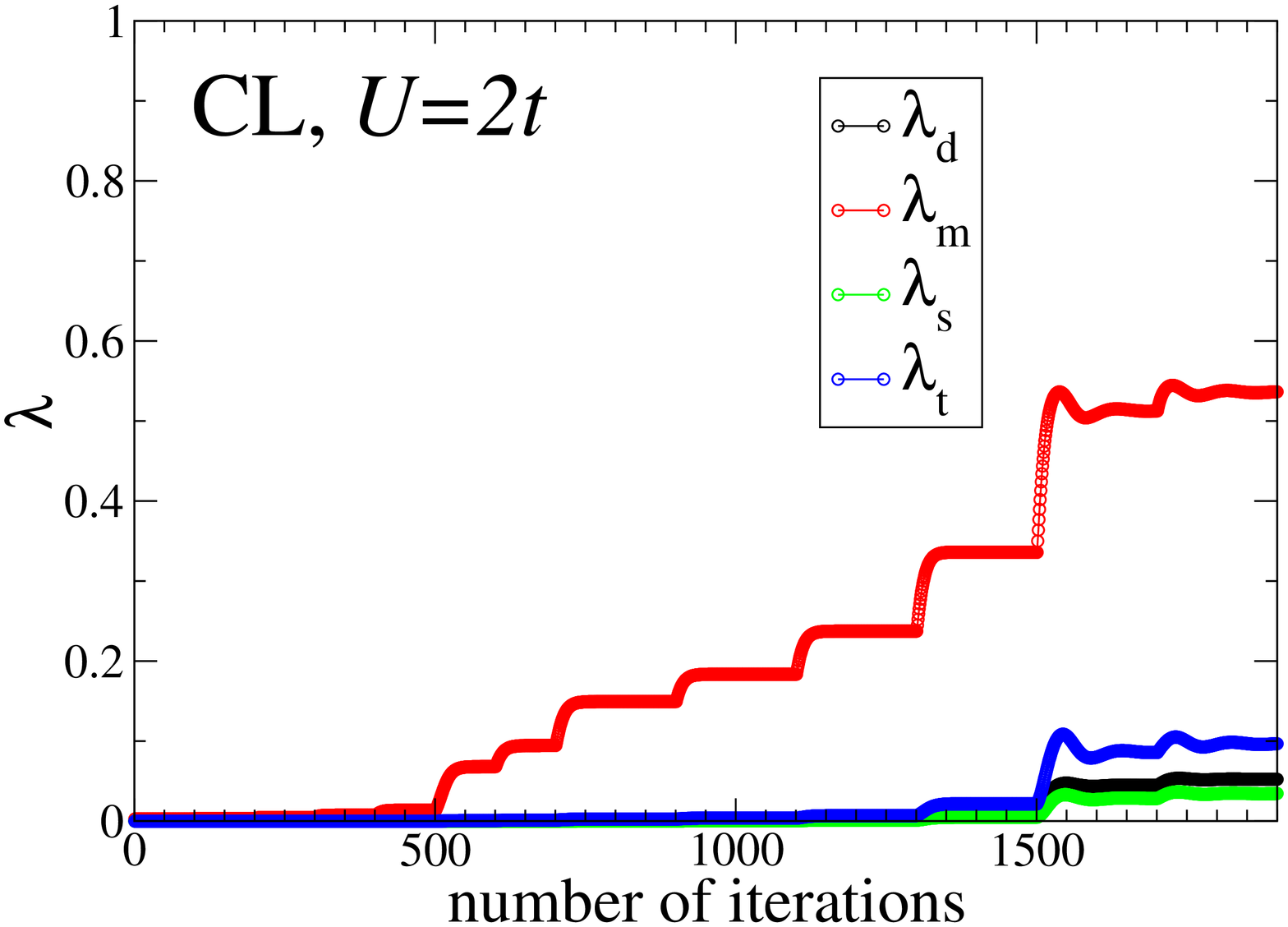}}
\centerline{
\includegraphics*[height=0.31\textheight,width=0.26\textwidth, angle=270, clip]{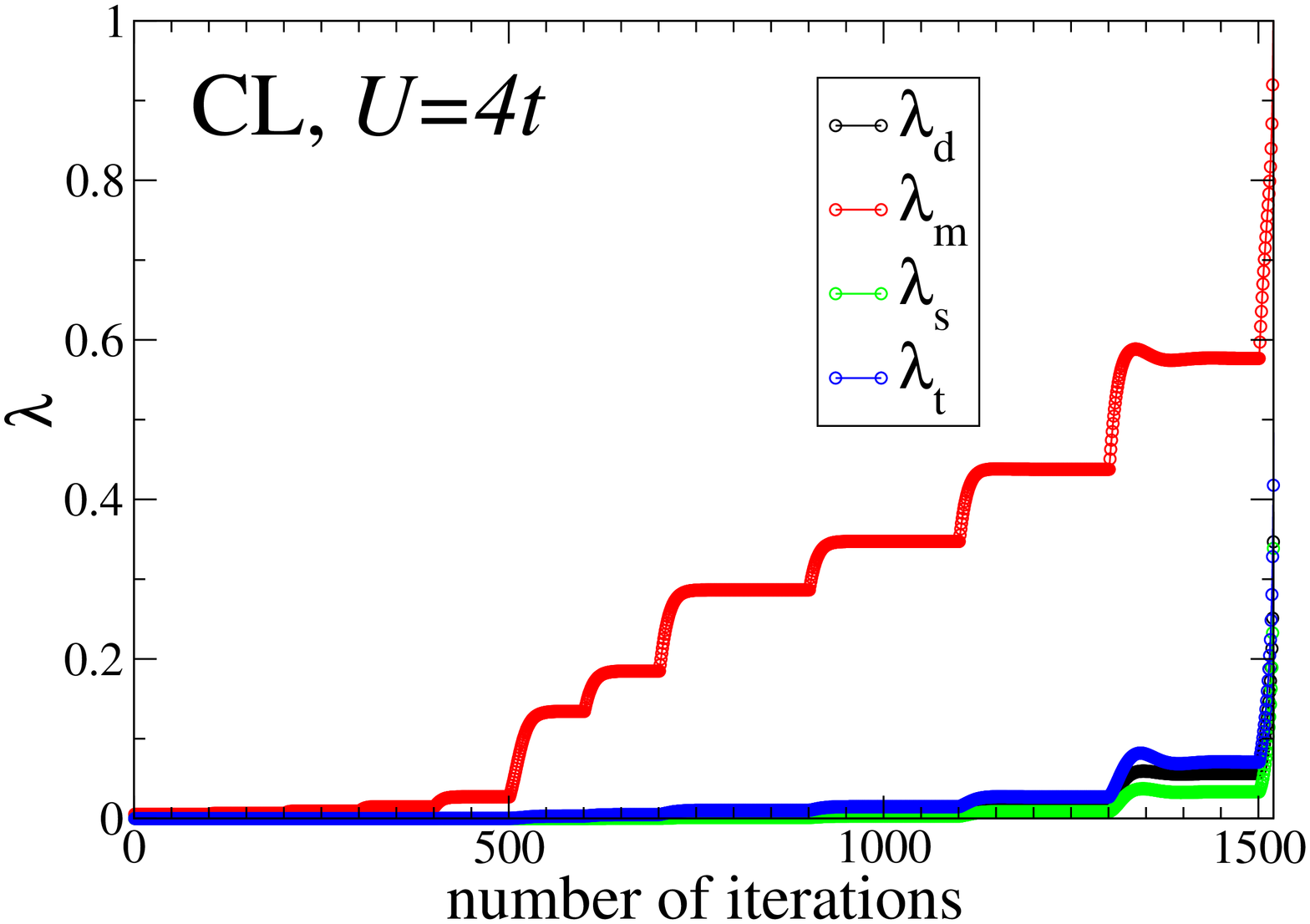}}
\centerline{
\includegraphics*[height=0.31\textheight,width=0.26\textwidth, angle=270, clip]{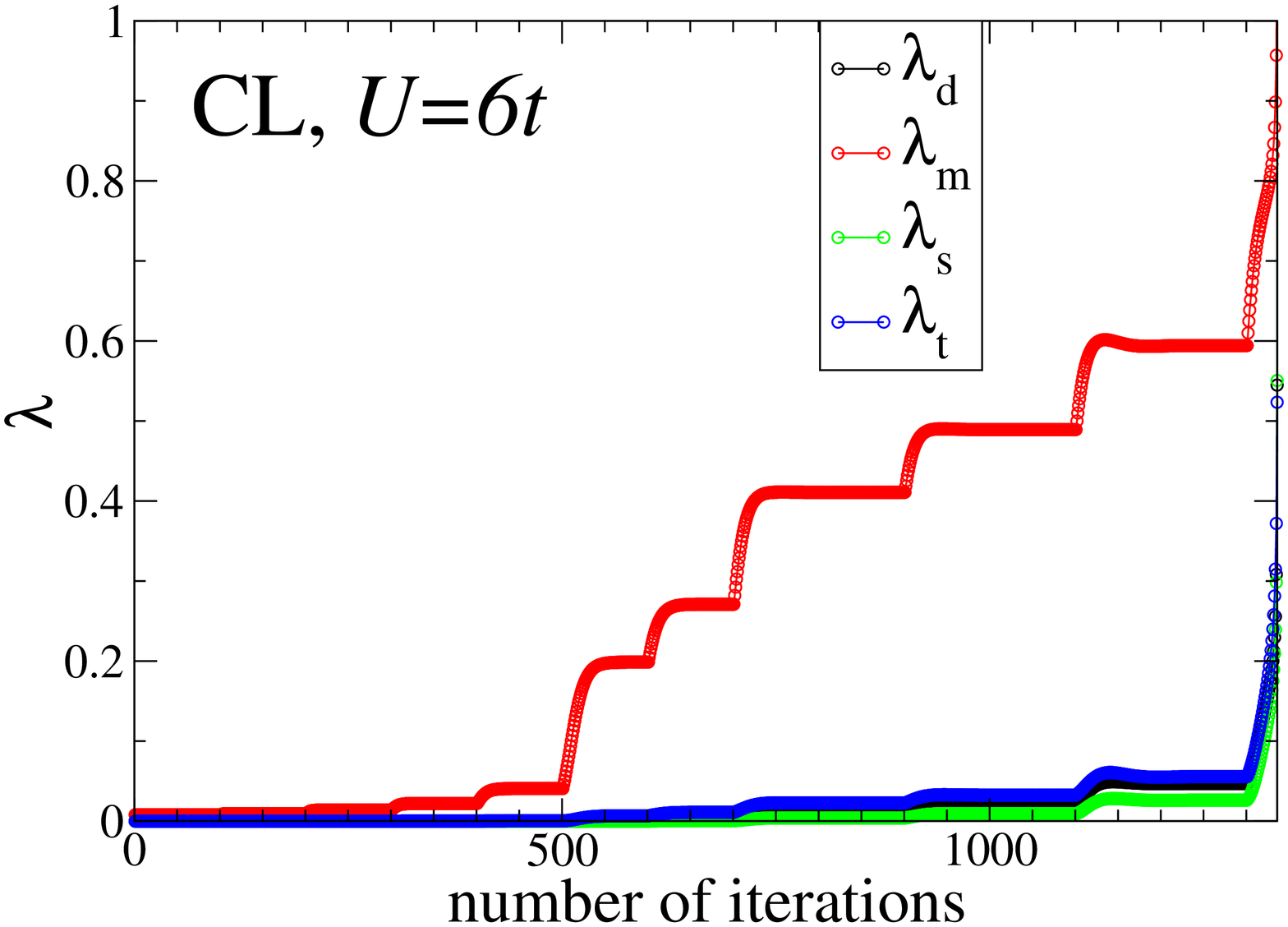}}

\caption{(Color online) The leading eigenvalues obtained by the continuous loading (CL) method. 
Symbolically we write the parquet equations as ${\bf f}_{parquet}(\Sigma,\Gamma)= 0$ and 
use them to define the auxiliary function as ${\bf g}=\nu {\bf f}_{parquet}+(1-\nu){\bf f}_{0}$, where 
${\bf f}_{0}$ is a function with a trivial solution. 
We choose ${\bf f}_{0}$ as a vector containing all the elements of $\Gamma - \Gamma_{0}$ and
$\Sigma - \Sigma_{0}$. The iteration method is used to solve the function ${\bf g}(\nu)$, 
instead of the ${\bf f}_{parquet}$. The solution of the ${\bf f}_{parquet}$ is recovered when $\nu=1$. 
A series of $\nu$ values are needed, which we denote as $\nu_{i}$. The function ${\bf g}(\nu_{i})$ is solved
by the simple iteration method with the initial conditions given by
the converged solution of the function ${\bf g}(\nu_{i-1})$.
For $U=2t$, we can push the value of $\nu$ to $1$ to
obtain the converged solution for $g(\nu=1)$, and the solution of
the parquet equations is recovered. However, the iteration procedure
diverges for the cases of $U=4t$ and $U=6t$; they diverge at $\nu=0.9999$
and $0.999$ respectively. These examples show that for the cases where
simple iteration method diverges, the continuous loading method may
not eliminate the divergence, even though the value of $\nu$ is pushed fairy close
to $1$.\label{continue_nosym}}
\end{figure}

In Fig.~\ref{continue_nosym_crossing}, we plot the crossing symmetry violation 
as a function of iteration steps with the continuous loading method and the 
same parameters as in Fig.~\ref{continue_nosym}. It is important to note that 
the homotopy function $g(\nu)$ does not respect the crossing symmetry except 
at $\nu=1$. Therefore, although the solution is converged, as long as $\nu \neq 1$, 
the crossing symmetry is violated.  The data for $U=2t, 4t$, and $6t$ are 
shown in the upper, the middle, and the lower panels respectively. The data
for $U=2t$ show a peak at the beginning of the iteration procedure
when $\nu$ is increased and gradually converges to a finite value. For
$\nu$ close to $1$, the data show a similar oscillatory behavior as
that from the simple iteration method. Similar behaviors are also
observed for $U=4t$ and $U=6t$, however, the iterations fail to
converge for $\nu$ close to $1$; the crossing symmetry is strongly
violated. Similar to that for the simple iteration method, at the
verge of the divergence, there are sharp increases of the violation of
crossing symmetry.

\begin{figure}
\centerline{
\includegraphics*[height=0.31\textheight,width=0.26\textwidth, angle=270, clip]{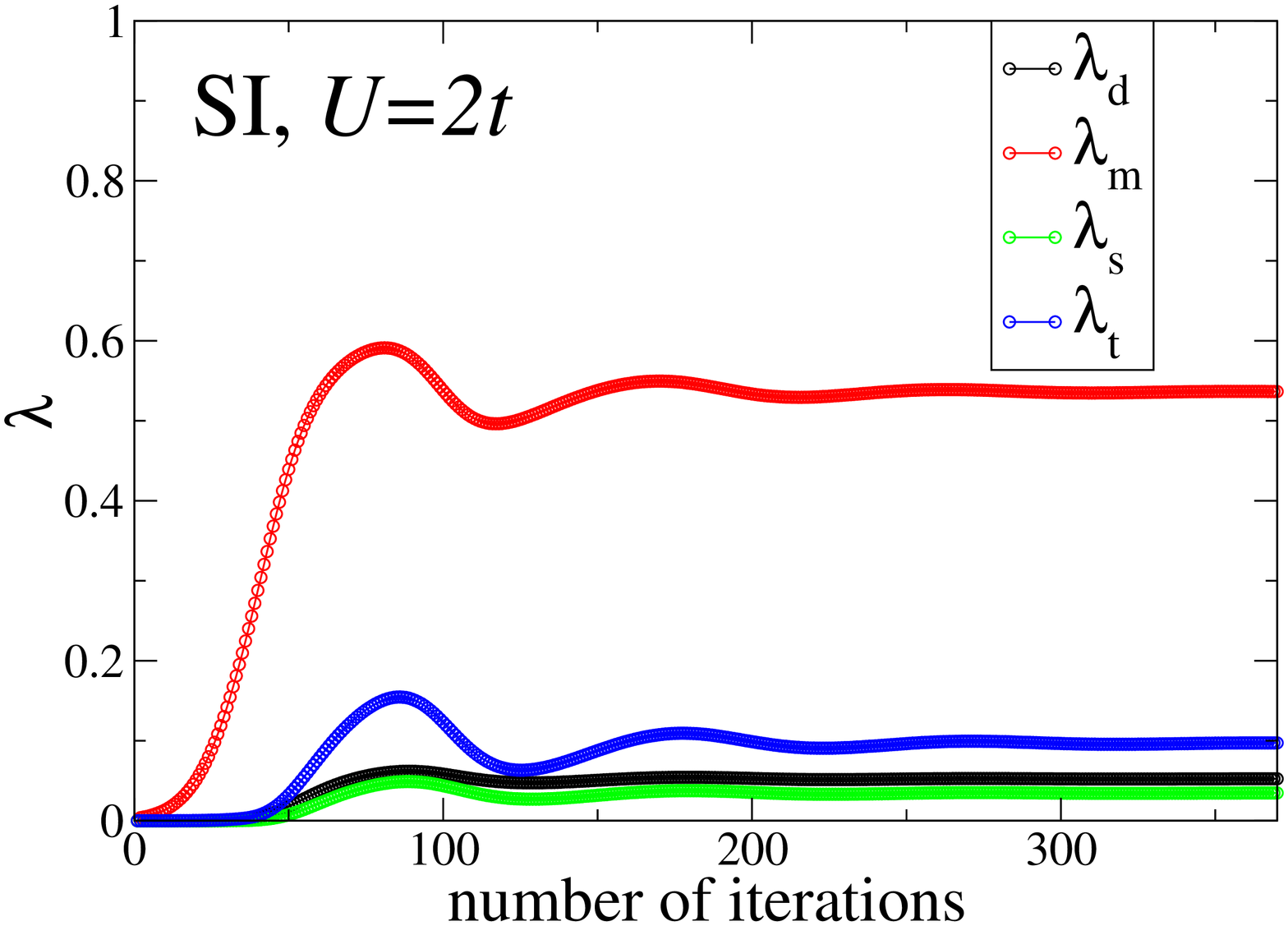}}
\centerline{
\includegraphics*[height=0.31\textheight,width=0.26\textwidth, angle=270, clip]{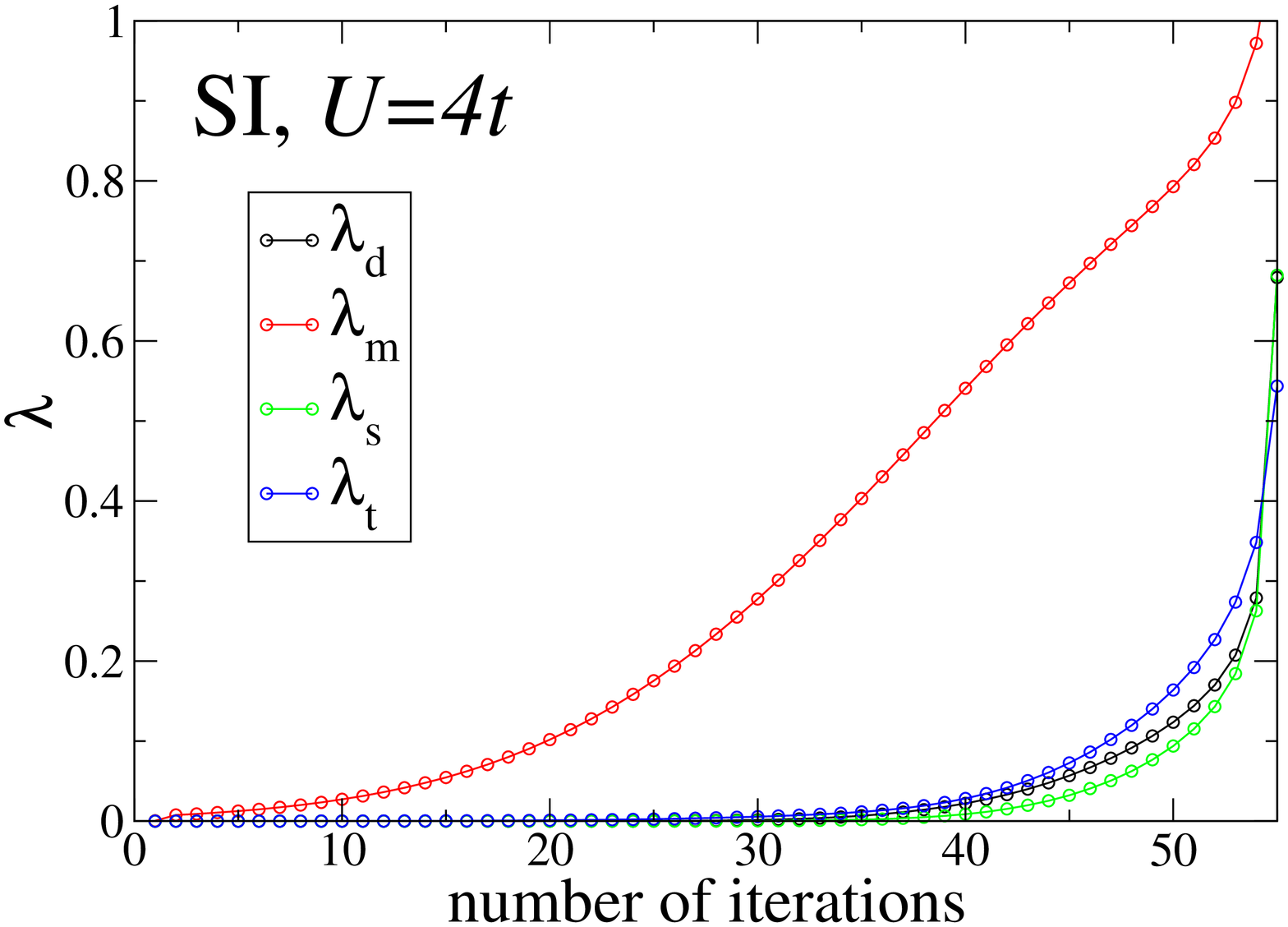}}
\centerline{
\includegraphics*[height=0.31\textheight,width=0.26\textwidth, angle=270, clip]{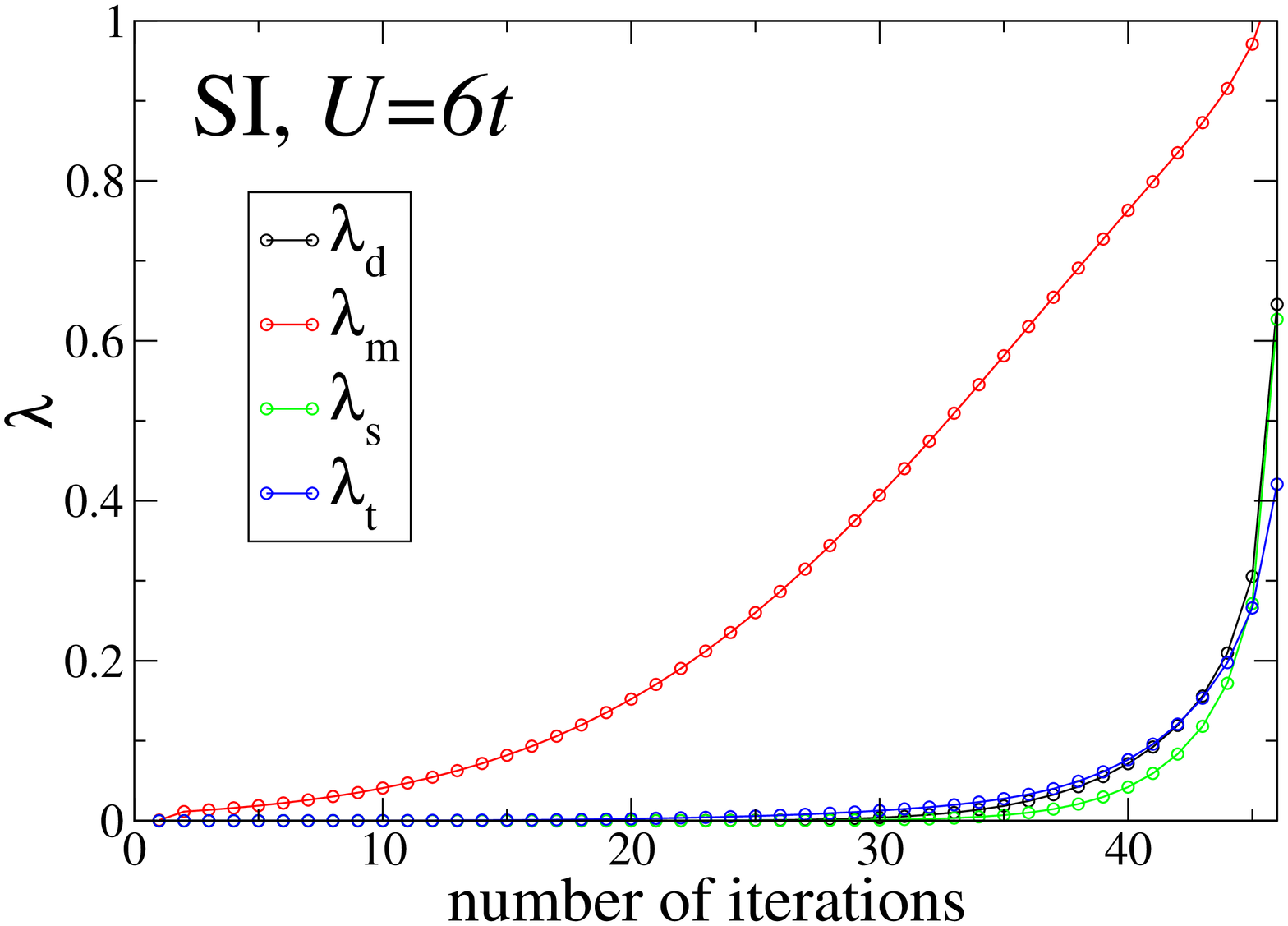}}
\caption{(Color online) The leading eigenvalues of various channels (density (d), magnetic (m), spin singlet (s), and spin
triplet (t)) as a function of the number of iteration
steps calculated with the simple iteration (SI) method. 
The calculations are for the half-filled Hubbard model on a $2
\times 2$ cluster at temperature $T=0.4t$. The initial condition for
the self-energy is set at $0+i320t$, and that for the irreducible
vertex is set at the bare Hubbard coupling. The damping factor
$\alpha$ is set at $0.1$. 
The solution is well converged for $U=2t$. However, divergence occurs for
$U=4t$ at the $55$th iteration step. For $U=6t$, divergence occurs at
the $46$th iteration step.  
\label{nosym}}
\end{figure}

\begin{figure}
\centerline{
\includegraphics*[height=0.31\textheight,width=0.26\textwidth, angle=270, clip]{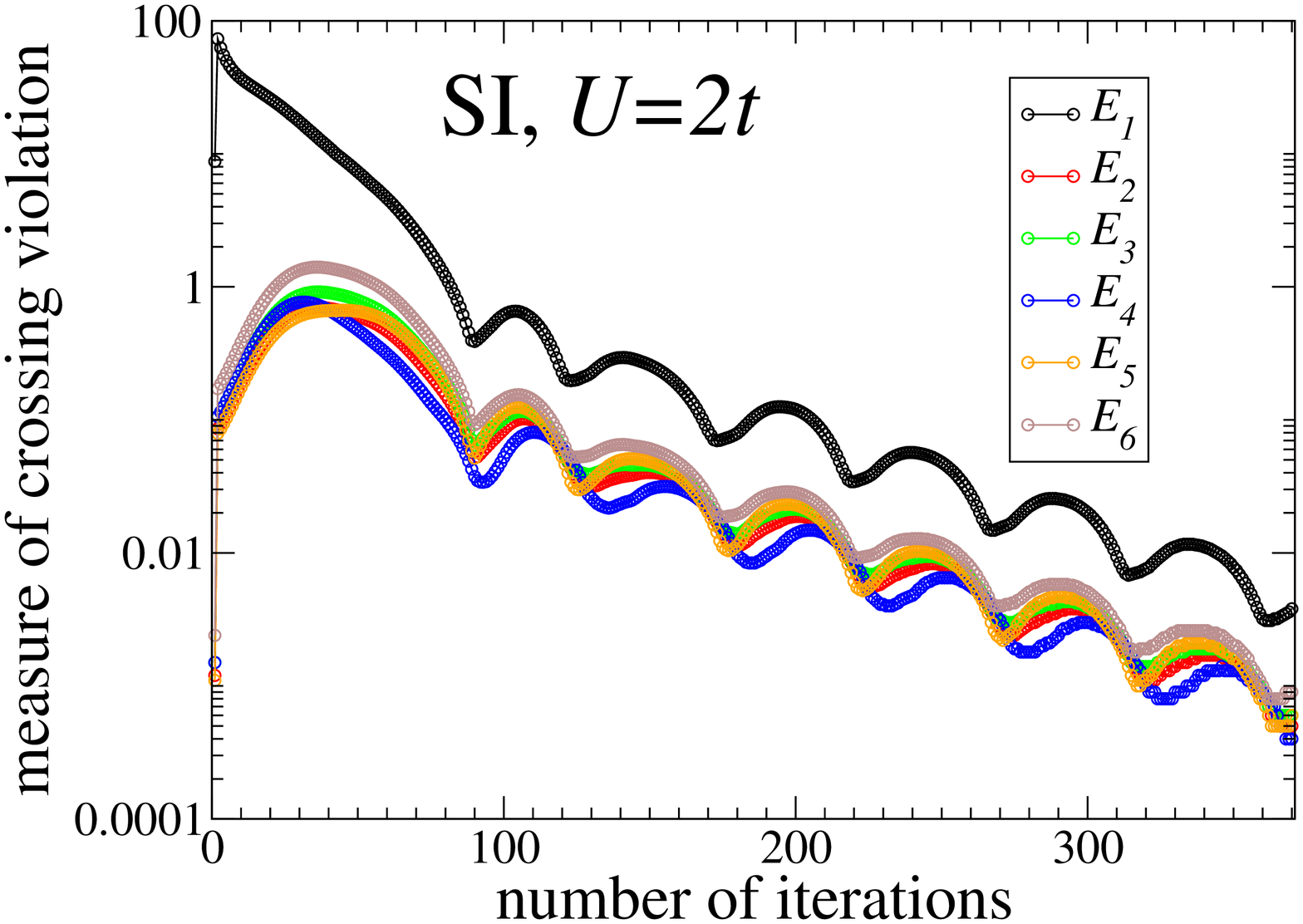}}
\centerline{
\includegraphics*[height=0.31\textheight,width=0.26\textwidth, angle=270, clip]{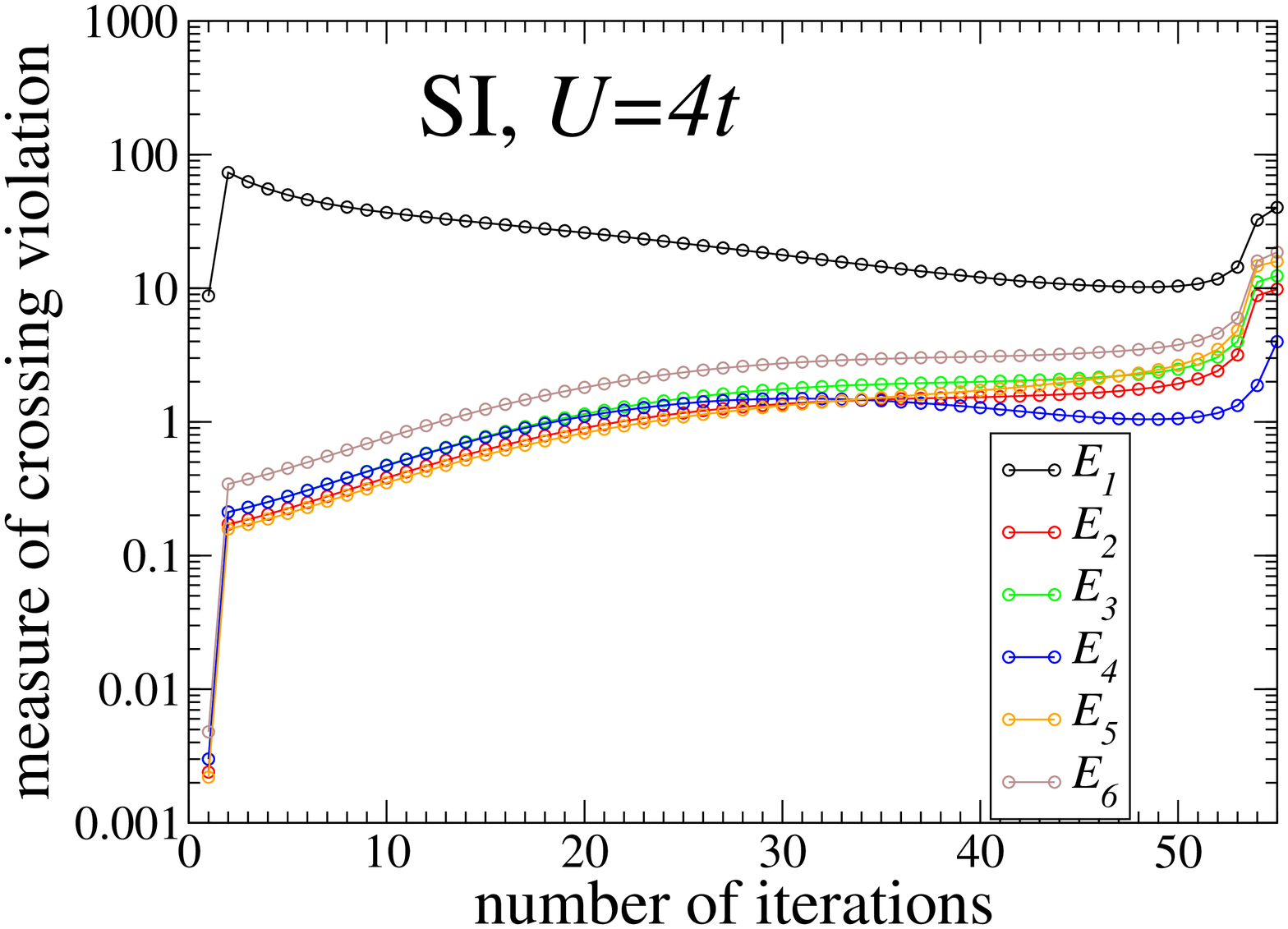}}
\centerline{
\includegraphics*[height=0.31\textheight,width=0.26\textwidth, angle=270, clip]{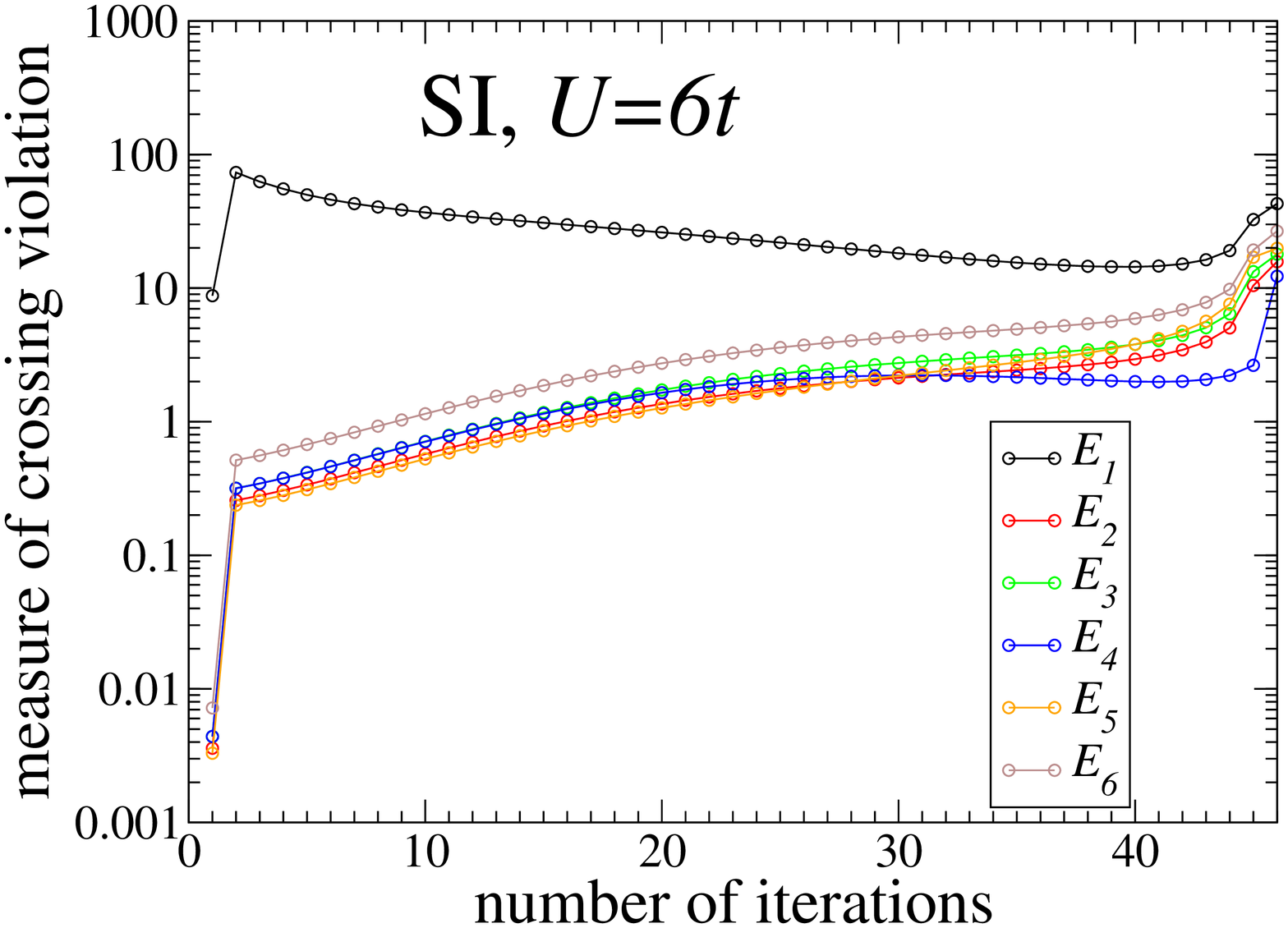}}
\caption{(Color online) Crossing symmetry violation, $E_{i}$, versus the number of iterations for the simple iteration 
(SI) method with the same parameters as Fig.~\ref{nosym}.
The six measures of crossing symmetry violation are defined as $E_{i} = |L_{i}-R_{i}| /
|L_{i}+R_{i}|$, where $i = 1,2, ... , 6$; $L_{i}$ and $R_{i}$ are defined
respectively as the left hand side and the right hand side of the
Eqs.~\ref{crossing_equalities1} -- \ref{crossing_equalities6}. 
The data for $U=2t$ shows an oscillatory but decreasing
trend. This is expected for the case where the iteration provides a
well converged solution. One should note that although the leading
eigenvalues seem to be converged, the crossing symmetry is not
perfectly constructed from the iteration.  For $U=4t$ and $U=6t$, the
iteration fails to provide converged solutions, and the crossing
symmetry is strongly violated. In particular, at the verge of the
divergence, there is a sharp increase of the violation of crossing
symmetry.
\label{nosym_crossing}}
\end{figure}

\begin{figure}
\centerline{
\includegraphics*[height=0.31\textheight,width=0.26\textwidth, angle=270, clip]{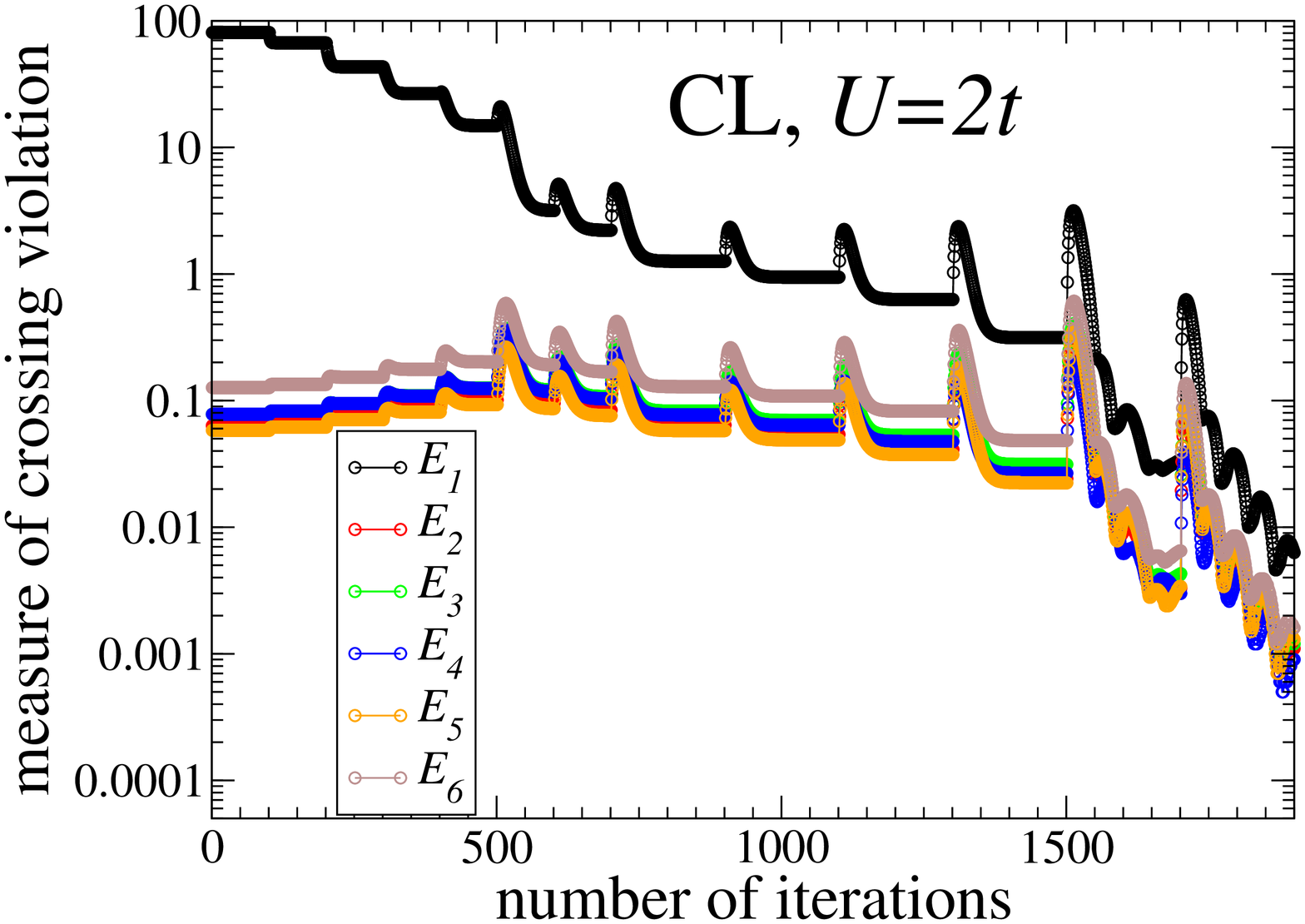}}
\centerline{
\includegraphics*[height=0.31\textheight,width=0.26\textwidth, angle=270, clip]{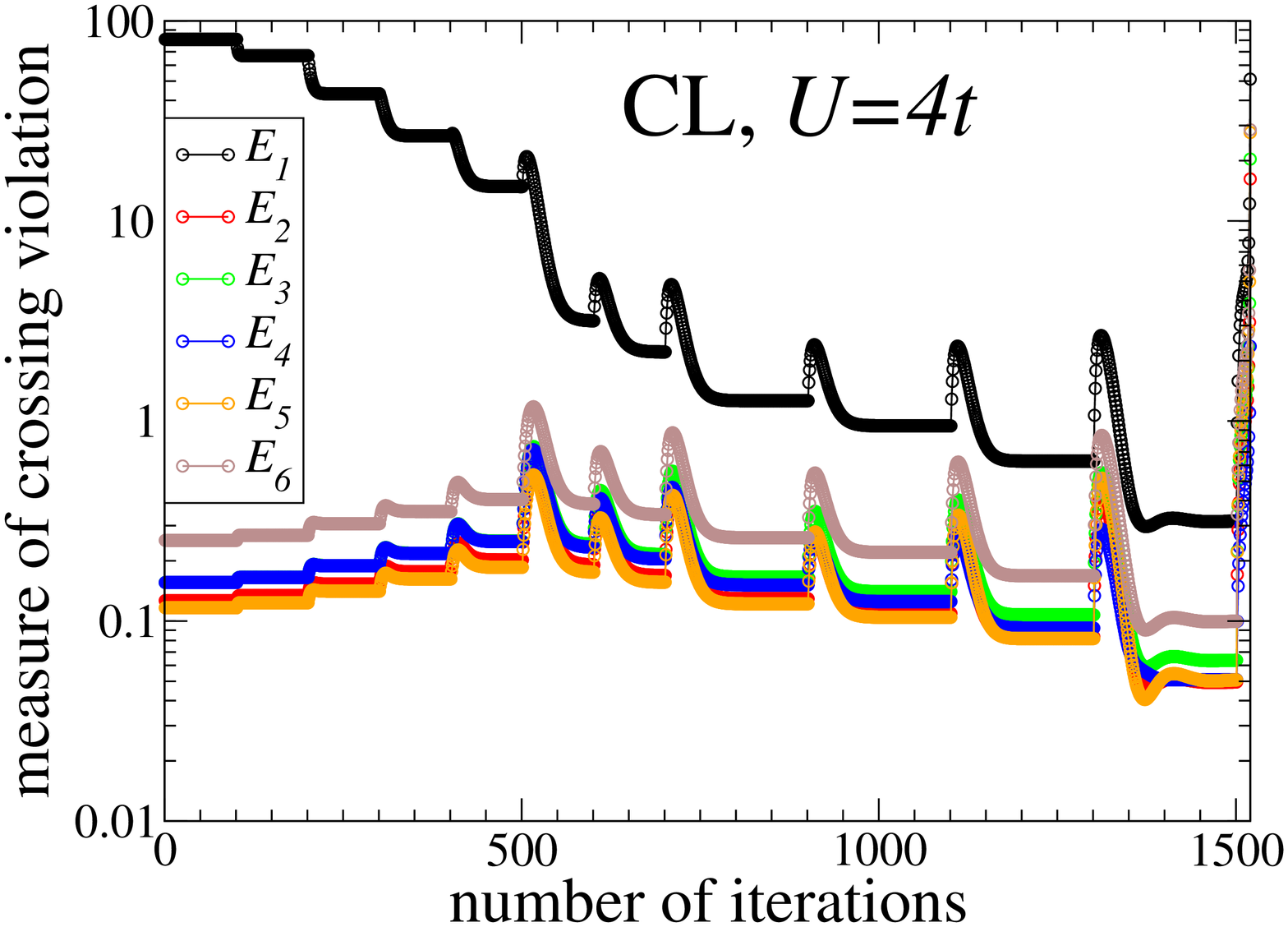}}
\centerline{
\includegraphics*[height=0.31\textheight,width=0.26\textwidth, angle=270, clip]{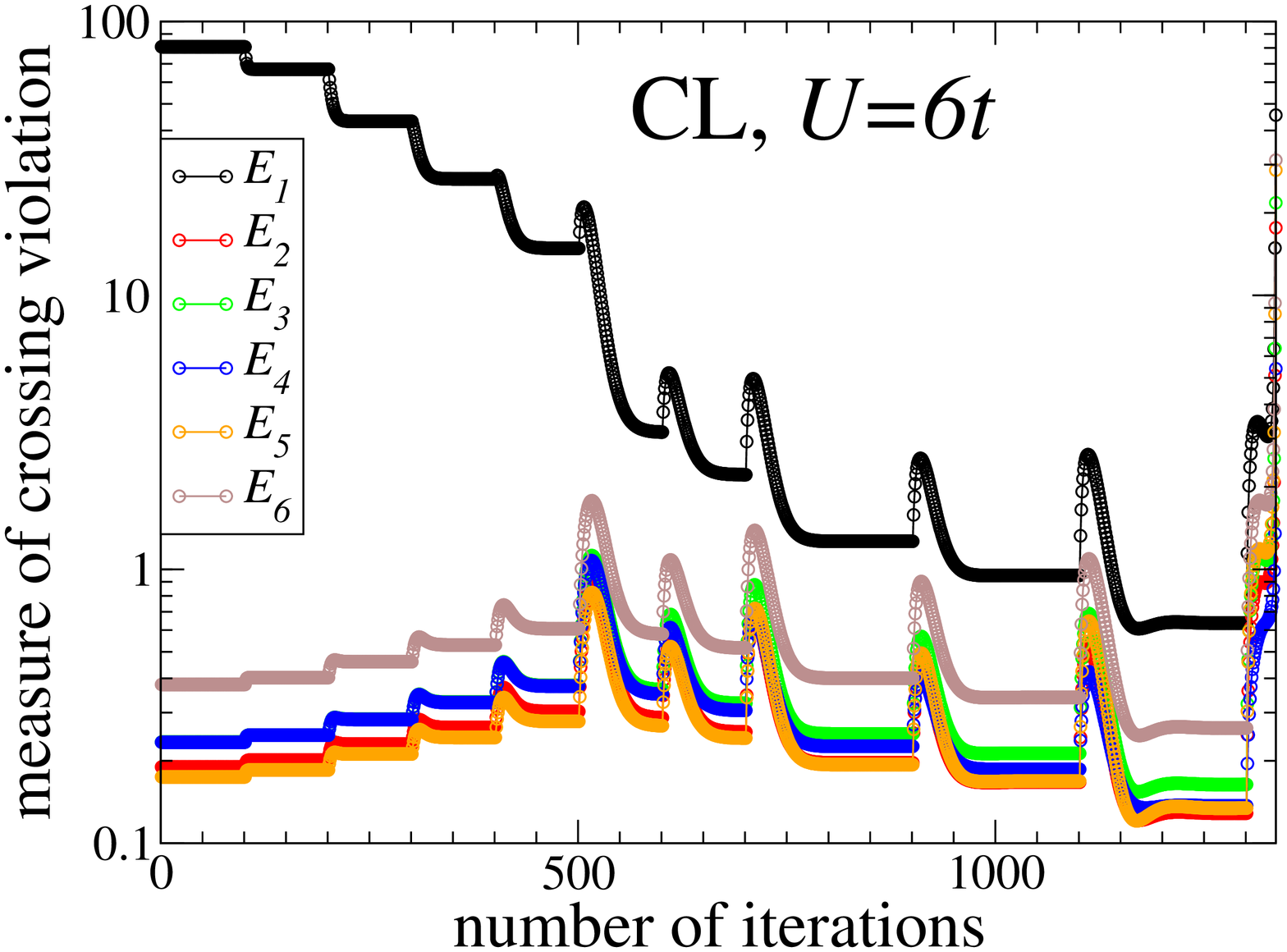}}
\caption{(Color online) Crossing symmetry violation, $E_{i}$, versus the number of iterations for the continuous loading (CL) 
method with the same parameters and the same definition of $E_{i}$ as in Fig.~\ref{nosym_crossing}.
The homotopy function $g(\nu)$  does not respect the crossing symmetry except at $\nu=1$. Therefore, although 
the solution may be converged for some values of $\nu$, as long as $\nu \neq 1$, the violation of crossing 
symmetry is non-zero. For $U=2t$ the crossing symmetry violations peak near the beginning of the iteration procedure 
where $\nu$ is small and gradually converge to a finite value when $\nu=1$.  Similar behaviors are also observed 
for $U=4t$ and $U=6t$; however, since the iteration fails to converge for $\nu$ close to $1$, the crossing symmetry
is strongly violated. Similar to that observed in the simple iteration method, at the verge of the divergence, 
there is  a sharp increase of the violation of crossing symmetry.
\label{continue_nosym_crossing}} 
\end{figure}

\section{Symmetry Restoration}
\subsection{Crossing Symmetry}
Although it does not seem to be easy to analyze all the causes of the
instability in the iteration, based on the discussion in the above
section, our conjecture is that one of the possible reasons for the instability is that
certain symmetries are violated in the course of the iteration
process. A possible strategy to improve the iteration scheme is to
impose those symmetries explicitly into the iteration process, so that
at each step of the iteration these symmetries are not violated.  The
full vertices $F$ obtained by the solution of the Bethe-Salpeter
equation cannot guarantee the crossing symmetry, unless the exact
solution is attained.  Therefore, so as to preserve the crossing
symmetry, the simplest method is to use the full vertices obtained by
solving the Bethe-Salpeter equation (that is the full vertices
obtained from the step 3 of the algorithm presented in the section III), and feed them back into the parquet
equation to reconstruct the crossing symmetric full vertices. 
Fig. \ref{algorithm_B} illustrates the flow diagram for solving the parquet equations with 
explicit restoration of the crossing symmetry in the full vertices.
Here is the algorithm which explicitly preserves the crossing
symmetry:

\begin{figure}
\centerline{
\includegraphics*[height=0.33\textheight,width=0.51\textwidth, viewport=0 92 692 510, clip]{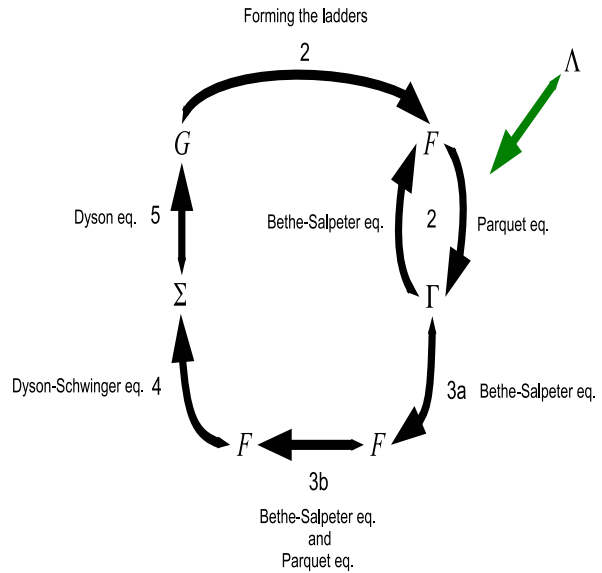}}
\caption{(Color online) Flow diagram of the algorithm for solving the parquet equations with crossing symmetry restoration. 
See the text for the description of each step.   The main difference compared to the previous algorithm is in 
the step 3b where the crossing symmetry is restored explicitly in the full vertex, $F$.  Because of this explicit
restoration of the crossing symmetry, in practice, the step 2 is only iterated for one time as the self-consistency 
is not required to generate the crossing symmetry.}
\label{algorithm_B}
\end{figure}

1. Set the initial conditions for the irreducible and full vertices, and the self-energy. 

2. Update the Green's functions and calculate the bare susceptibility, $\chi_0$. 
Solve the parquet and the Bethe-Salpeter equations for 
the irreducible vertices, $\Gamma$. Simple iteration is used until the convergence criteria are met for the
irreducible vertices.

Since we will restore the crossing symmetry of the full vertices in the step 3b, we find that it is not necessary 
to attain the self-consistency for step 2. In practice, we iterate the parquet equations for one time only. 
The next step is to use the irreducible vertices obtained from the parquet equations to construct the full vertices.

3a. Solve the Bethe-Salpeter equation to obtain the full vertices, $F$, 
    using the irreducible vertices from the previous step. This is
    executed exactly by calling the LAPACK routines for the inverse of
    the matrices. \cite{lapack}

3b. Use the new irreducible vertices obtained in step 2 and the full
    vertices obtained in step 3a to form the vertex ladders.
    Construct the full vertices from Eqs.~\ref{SD_BSPH_EQ} and
    \ref{SD_BSPP_EQ} using the vertex ladders. Following these steps, the crossing symmetry
    is restored in the full vertex $F$.

With the full vertices obtained, we can update the self-energy.

4. Solve the Dyson-Schwinger equation to obtain the self-energy from
   the full vertices. Simple iteration is used until the
   convergence criteria are met for the self-energy.

5. Solve the Dyson equation for the fully dressed Green function from
   the self-energy. 

This completes the iteration loop, and the procedure is repeated from
step 2 until the criteria of convergence are met for both the
self-energy and the irreducible vertices.

\begin{figure}
\centerline{
\includegraphics*[height=0.31\textheight,width=0.26\textwidth, angle=270, clip]{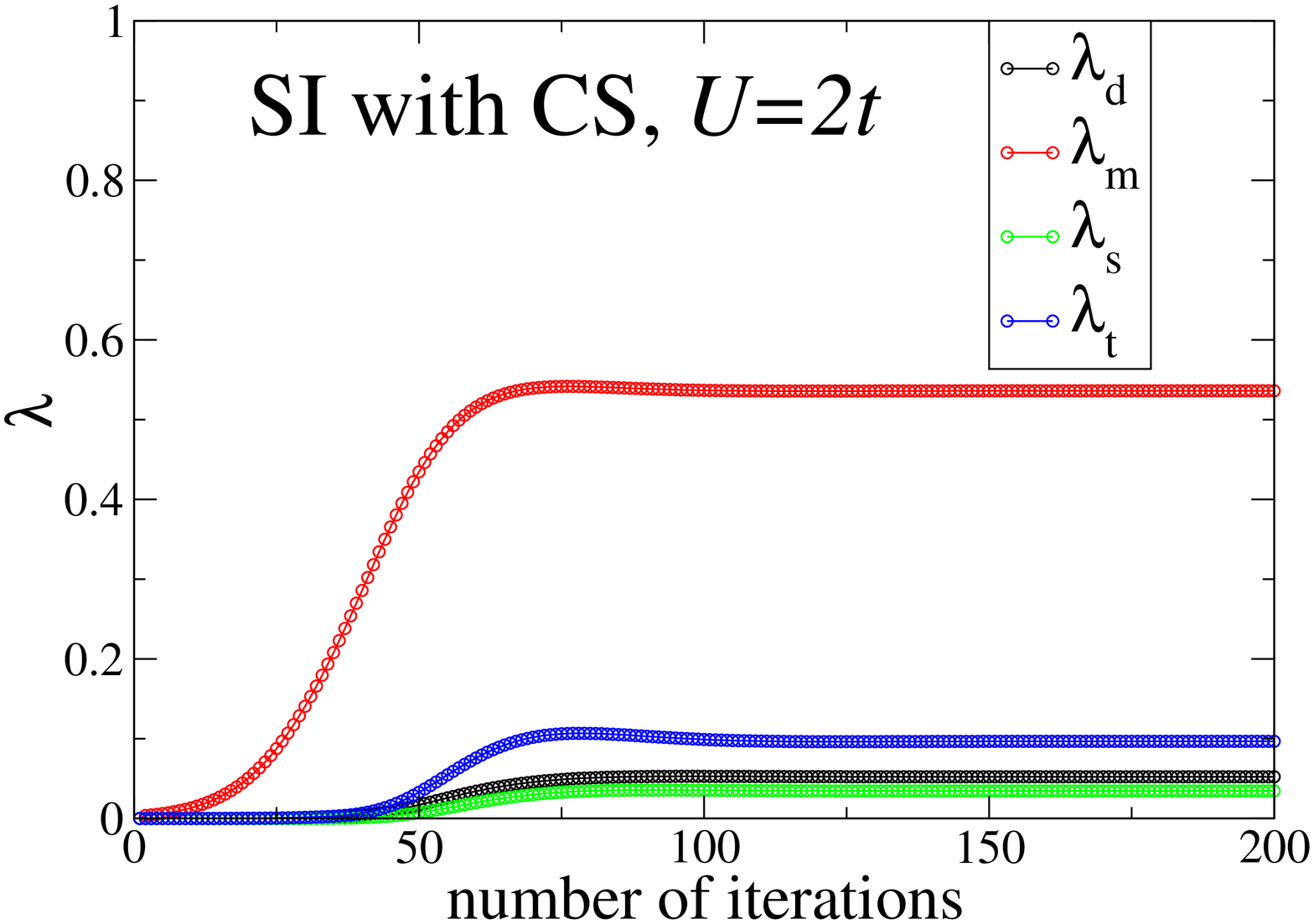}}
\centerline{
\includegraphics*[height=0.31\textheight,width=0.26\textwidth, angle=270, clip]{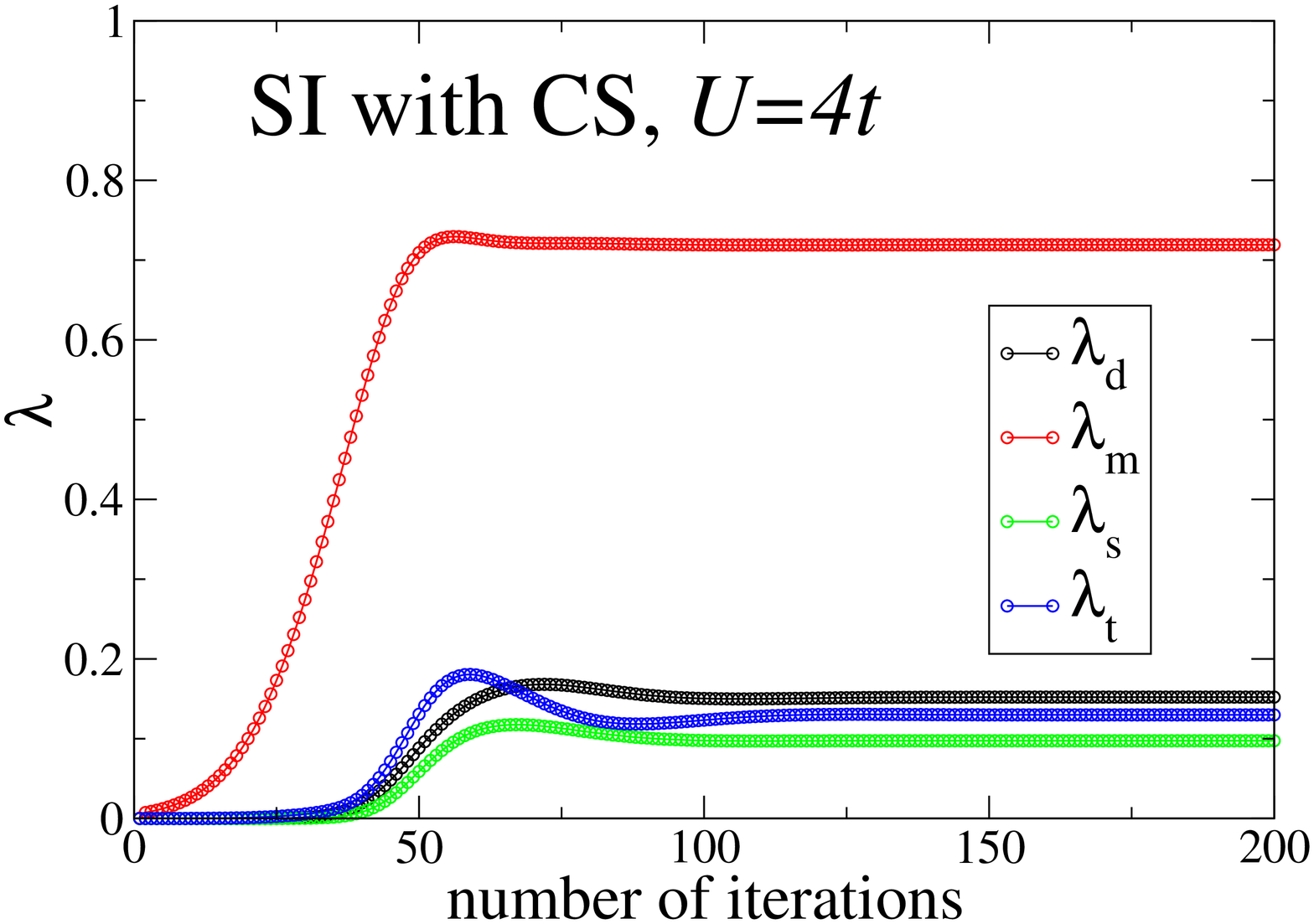}}
\centerline{
\includegraphics*[height=0.31\textheight,width=0.26\textwidth, angle=270, clip]{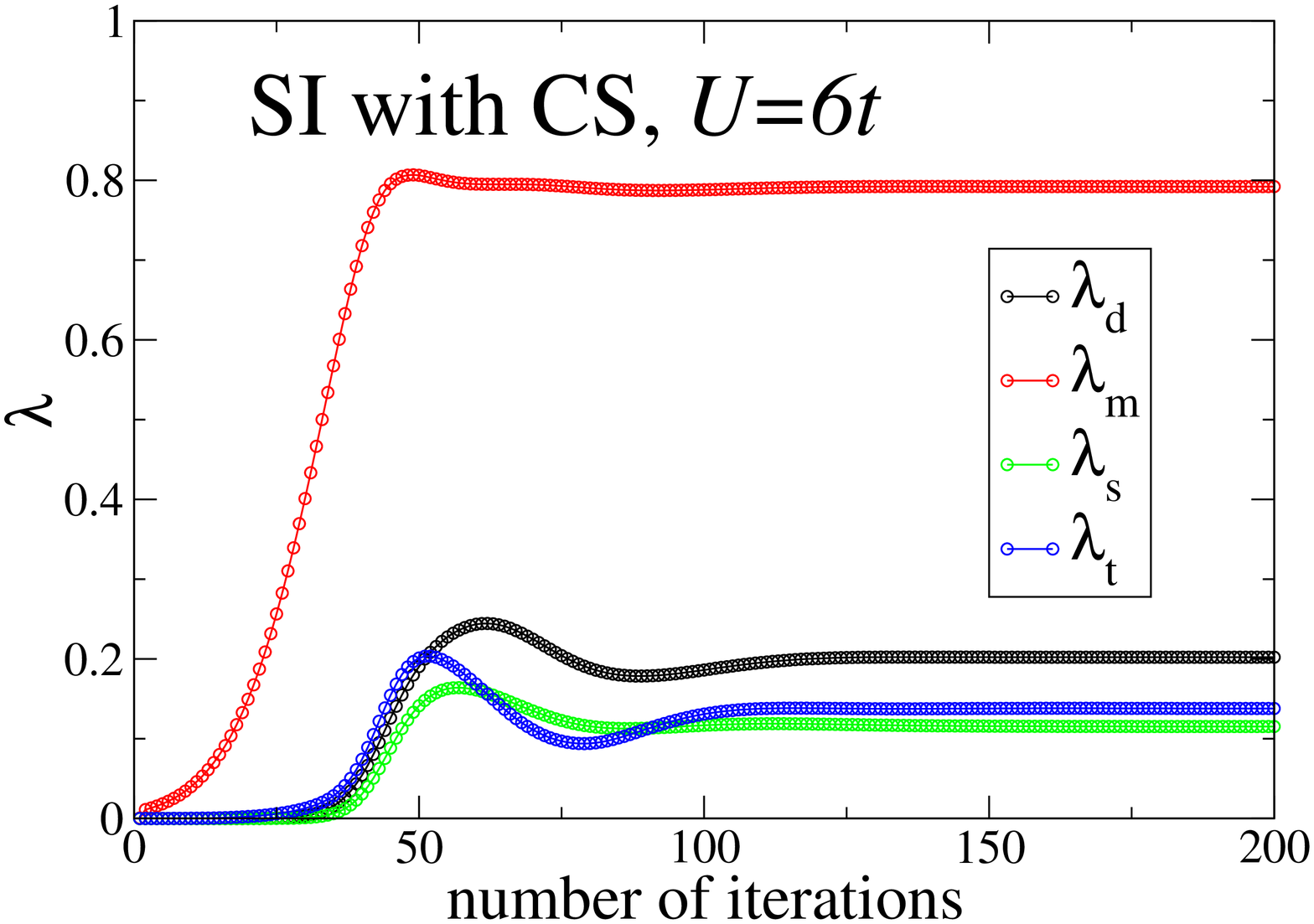}}
\caption{(Color online) The leading eigenvalues of various channels (density (d), magnetic (m), spin singlet (s), 
and spin triplet (t)) versus the number of iterations 
with the simple interaction (SI) method with crossing symmetry (CS). 
The parameters used are the same as the data shown in
Fig.~\ref{nosym}.  The only difference is that the crossing symmetry in
the full vertex, $F$, is explicitly restored at each step of the
iteration. This is easily achieved by constructing the full vertex
directly from Eqs.~\ref{SD_BSPH_EQ} and \ref{SD_BSPP_EQ}. 
The simple iteration scheme without crossing
symmetry fails for the case of $U=4$ and $6$. With the crossing
symmetry explicitly restored, converged solutions are obtained.
\label{sym}}
\end{figure}

The main difference between the current algorithm and the previous
algorithm we present in Section III is in step 3 where the full
vertices are constructed. In the previous algorithm the Bethe-Salpeter
equation is solved many times to attain convergence. When the
absolute convergence is attained, the crossing symmetry will be
satisfied. In the current algorithm, we just explicitly solve the
Bethe-Salpeter equation in step 3a to refresh the full
vertices. Once we obtain the full vertices, in step 3b, we
construct the new vertex ladders and the new full vertices from the
vertex ladders using the Bethe-Salpeter equation. 
By doing so, the crossing symmetry of the full vertices will
be satisfied; see Eqs.~\ref{F_d}, \ref{F_m}, \ref{F_s}, and \ref{F_t}.
In Fig.~\ref{sym} we show the leading eigenvalues
using the same set of parameters used in 
Fig.~\ref{nosym}. While the simple iteration scheme without crossing
symmetry fails to converge for the case of $U=4$ and $6$, it provides a converged solution 
when the crossing symmetry is explicitly restored.

\begin{figure}[h!]
\centerline{
\includegraphics*[height=0.31\textheight,width=0.26\textwidth, angle=270, clip]{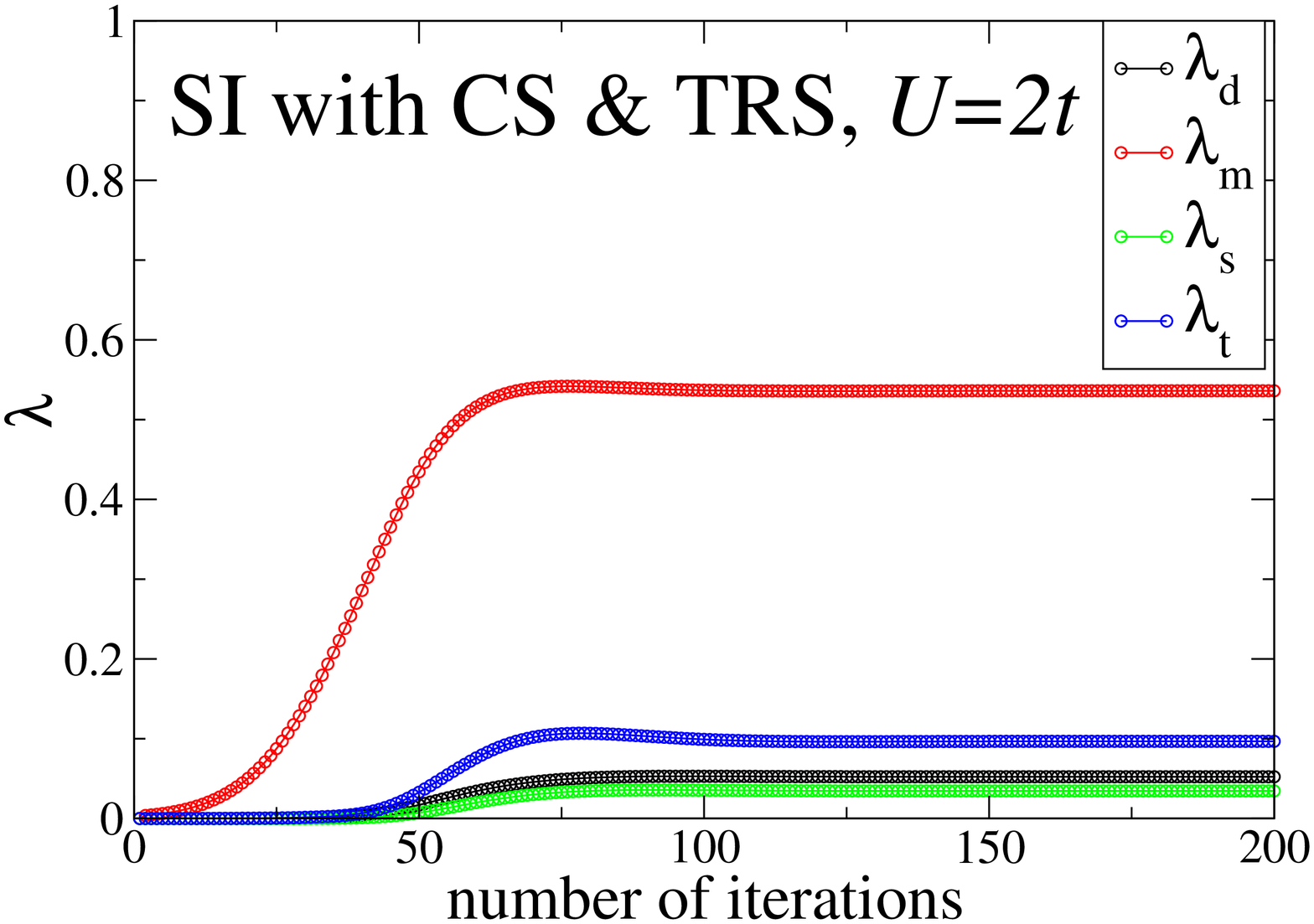}}
\centerline{
\includegraphics*[height=0.31\textheight,width=0.26\textwidth, angle=270, clip]{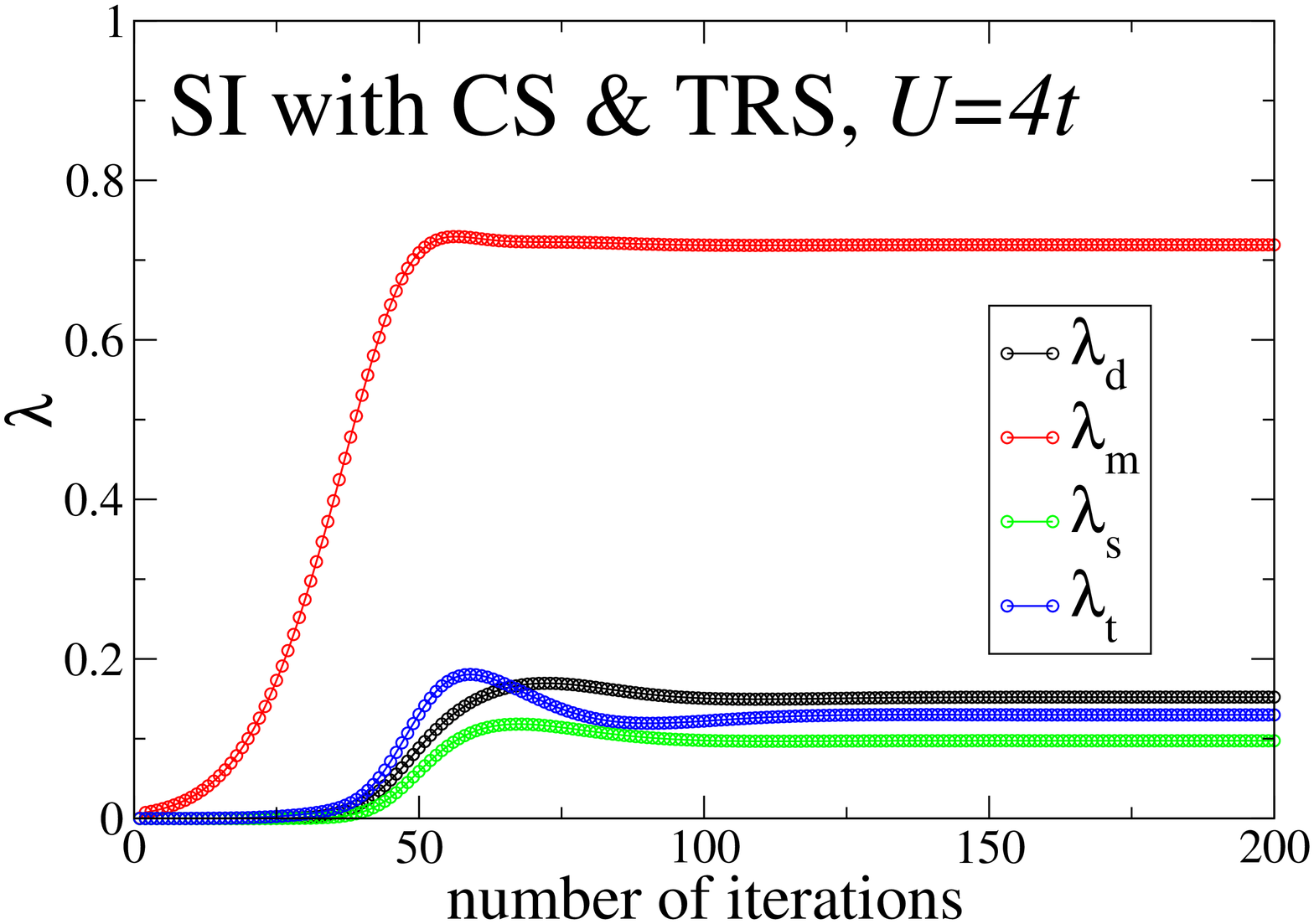}}
\centerline{
\includegraphics*[height=0.31\textheight,width=0.26\textwidth, angle=270, clip]{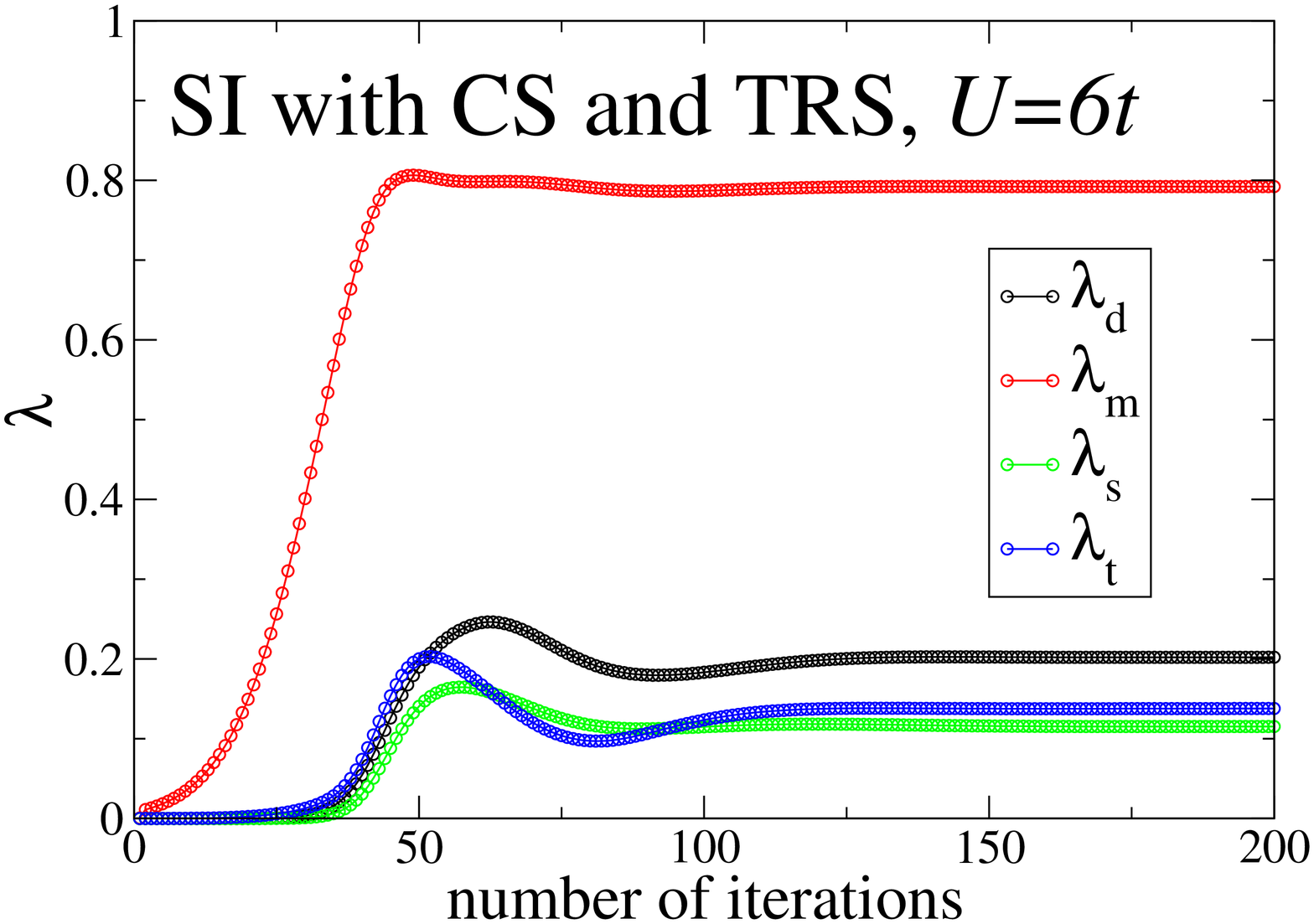}}
\caption{(Color online) The leading eigenvalues of various channels: density (d),
magnetic (m), spin singlet (s), and spin triplet (t), as a
function of the iteration. 
The parameters used are the same as the data shown in
Fig.~\ref{nosym}.  Two symmetries are explicitly restored at each
step of the iteration: the crossing symmetry (CS) in the full vertex,
$F$, and the time-reversal symmetry (TRS) for the self-energy, $\Sigma$, and both the irreducible vertex,
$\Gamma$, and the full vertex, $F$, (that is $F(Q)_{P,P^{'}} =F(Q)_{P^{'},P}$ and similarly for $\Gamma$).
Notice there is no substantial gain in the convergent rate compared to the
case with only the crossing symmetry being restored.
\label{fullsym}}
\end{figure}

\subsection{Time-reversal and Point Group Symmetry}

Besides imposing the crossing symmetry on the full vertices, some of the internal
symmetries can also be imposed on the irreducible vertices and the self-energy without
much computational overhead.  We illustrate the time-reversal symmetry
in the self-energy
\begin{equation}
\Sigma(\mathbf{k},i\omega) = \Sigma^{*}(\mathbf{k},-i\omega)
\end{equation}
and the vertices (spatial reflection symmetry and parity invariance are assumed), 
\begin{equation}
F(Q)_{P,P^{'}} = F(Q)_{P^{'},P}.
\end{equation}

As these symmetry operations do not mix vertices across different 
values of $Q$, and providing that the data is distributed with 
one or more $Q$ at each node, the time reversal symmetry of the vertices 
can be imposed without invoking communications across different nodes.
Therefore, enforcing time-reversal symmetry will only cause a very minor 
computational overhead.

Other symmetries, such as the point group symmetry for the square lattice 
can be rather cumbersome.  An expensive scheme involving heavy internode 
communication would be required to impose the complete set of point group 
symmetries. However, we may impose an important subset of the operations 
$R_{\alpha}$ for which $R_{\alpha}(Q)=Q$ without expensive communications. In 
these cases, the vertices may be symmetrized by performing the sum
\begin{equation}
 F(Q)_{P,P^{'}} = \frac{1}{N_{R_\alpha(Q)=Q}}\sum_{R_\alpha(Q)=Q} F(Q)_{R_\alpha(P),R_\alpha(P^{'})},
\end{equation}
where $N_{R_\alpha(Q)=Q}$ is the number of elements in this subset of operations.
For general $Q$ in the cluster Brillouin zone there would be no $\alpha$ such
that $R_\alpha(Q)=Q$ apart from the identity. However, for the points of high 
symmetry, $R_\alpha(Q)=Q$ for all $\alpha$. Generally, the instabilities first 
occur here, so imposing the point group symmetries at
these $Q$ values should have the greatest impact.

In Fig.~\ref{fullsym} we show the leading eigenvalues of various channels when both
crossing and time-reversal symmetries are imposed for
the same set of parameters being used from the data in Fig.~\ref{nosym} and \ref{sym}.
Spatial reflection symmetry, parity invariance and spin rotation invariance are
assumed as appropriate for the two-dimensional Hubbard model at non-zero temperature. 
We can see that there is only very marginal improvement for the convergence compared to 
the results without explicitly restoring the time-reversal symmetry (see 
Fig.~\ref{sym}).  We also use the scheme described above to partially 
impose the point group symmetries.  However, these symmetries resulted
in no additional improvements and therefore no results are shown.

\section{Leading Eigenvalue of the Antiferromagnetic Channel}

\begin{figure}
\centerline{
\includegraphics*[height=0.32\textheight,width=0.39\textwidth, angle=270, 
clip]{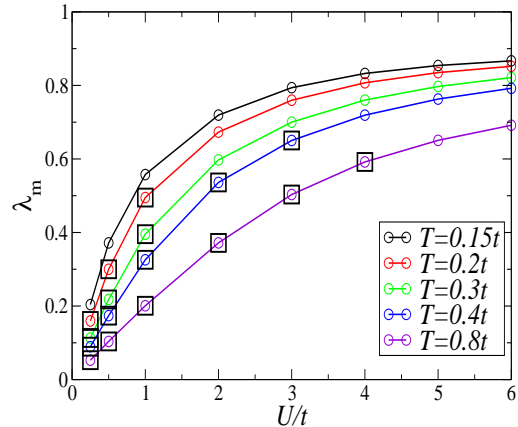}}
\caption{(Color online) The leading eigenvalues of the antiferromagnetic magnetic channel, $\lambda_{m}$,  
as a function of the coupling, $U$, calculated with the simple iteration method.
Different curves correspond to different temperatures. The data points 
enclosed in a black square correspond to the cases where the simple iteration 
without any symmetry restoration can provide a converged solution.
\label{eigen_ref}}
\end{figure}

With the improved scheme proposed in this paper, we are able to
explore a wider range of temperature and coupling strength for the
half-filled Hubbard model.  In Fig.~\ref{eigen_ref} we show the
leading eigenvalue for the most singular channel, 
the antiferromagnetic channel, $\lambda_{m}$, as a function of $U$ 
for a range of temperatures as low
as $T=0.15t$.  The data points enclosed in a black square correspond to the cases 
where the simple iteration without any symmetry restoration provides a converged solution.
For all temperatures, $\lambda_{m}$ increase sharply at weak coupling ($U \sim 2t$), and they tend to
saturate at strong coupling ($U \sim 6t$). They are most sensitive to
temperature at the intermediate coupling ($2t < U < 4t$).
We emphasize that convergency  is not possible
without the improved scheme, unless a large number of iterations are used 
to attain the crossing symmetry.

\section{Summary and Discussion}

We present improvements of numerical implementations for solving the 
parquet equations for the Hubbard model. The main strategy is to enforce 
the symmetries in the iteration process. The most prominent advantage of 
the parquet formulation, compared to most of the other approaches, is that 
the crossing symmetry is exactly fulfilled. However, in general, it is 
true only if the exact solution is found. With the simple iteration method, 
the crossing symmetry is strongly violated prior an instability, suggesting that the 
instability is due to these symmetry violations.

The continuous loading or homotopy method does not 
improve convergence significantly beyond the simple iteration method.  We note that the solutions of 
the continuous loading function do not preserve crossing symmetry. This may partly explain why the 
continuous loading method does not provide significant improvement over simple iteration. 

We present a simple method to enforce the crossing symmetry at each step of the 
iteration which does not substantially increase the computational cost. The addition
of these symmetry constraints can greatly improve the stability of the 
calculation, so that a wider range of parameters can be explored by the 
parquet formulation.  Along this line of thought, one can expect that 
the stability may be further improved if other symmetries are also imposed, 
the obvious ones being time-reversal and point group. However, these 
additional symmetries did not improve the stability significantly beyond 
that obtained with crossing symmetry alone.

The parquet formulation 
still remains as one of the best approaches for calculating the two-particle vertex 
functions in a self-consistent manner. At present, solving the parquet 
equations for a large lattice size is still a very challenging task; however, 
with the continuous advances of computational facilities, it should become
more feasible in the foreseeable future.  A promising direction, which 
allows immediate application of the parquet formulation, is to incorporate 
it as part of the multi-scale many-body approach.\cite{Jarrell-etal-2007,Slezak-etal-2009}

\begin{acknowledgments}
We would like to acknowledge very useful discussions with Karen Tomko, and 
we thank Peter Reis for his careful reading of the manuscript.
This work was supported in part by the DOE SciDAC grant DE-FC02-06ER25792
(KMT, HF, SYZ, and MJ) and the U.S. National Science Foundation LA-SiGMA 
grant EPS-1003897 (JR, JM, and MJ). Supercomputer support was provided by the NSF  
TeraGrid under grant number TG-DMR100007. This research also used resources of the National Center 
for Computational Sciences at Oak Ridge National Laboratory, which is supported by the Office of 
Science of the U.S. Department of Energy under Contract No.\ DE-AC05-00OR22725.

\end{acknowledgments}

\appendix


\section{Parallel Implementation with Latency Hiding} 
This appendix describes a highly effective implementation of the symmetry-enforcing 
variant of the parquet formulation described earlier in the paper. The communication 
bottleneck in this implementation is the expensive tensor rotations required to 
rotate the vertex ladders (Eqs.~\ref{eq:Phi_ladder} and \ref{eq:Psi_ladder}) between 
the forms used in the Bethe-Salpeter equation, Eqs.~\ref{SD_BSPH_EQ} and \ref{SD_BSPP_EQ}, 
to those used in the parquet equations, Eqs.~\ref{Gamma_d} -- \ref{Gamma_t}.  
If we distribute the equations between the processes executing on the compute nodes of a 
parallel machine using the transfer momenta $Q$, the tensor rotations are done with an expensive all-to-all 
communication among those processes, in which every node needs to communicate with 
all the other nodes. The MPI implementation of the all-to-all communication\cite{mpi} is a 
collective operation that is blocking, i.e., each process has to wait until the message has 
been sent out. The key aspect of our implementation is the decomposition of the required 
communication so that non-blocking communication primitives can be effectively utilized. 
The non-blocking communication enables latency hiding by
overlapping computations and communications.

Four different forms of tensor rotations are required:
\begin{eqnarray}
{\rm rotation \: 1:} \ \Phi(Q)_{P,P^{\prime}} & \longleftarrow & \Phi(P^{\prime}-P)_{P,P+Q} \label{rot1} \\ 
{\rm rotation \: 2:} \ \Phi(Q)_{P,P^{\prime}} & \longleftarrow & \Phi(P^{\prime}-P)_{-P^{\prime},P+Q} \label{rot2} \\
{\rm rotation \: 3:} \ \Phi(Q)_{P,P^{\prime}} & \longleftarrow & \Phi(P+P^{\prime}+Q)_{-P^{\prime},-P} \label{rot3} \\
{\rm rotation \: 4:} \ \Phi(Q)_{P,P^{\prime}} & \longleftarrow & \Phi(P+P^{\prime}+Q)_{-P-Q,-P}. \label{rot4}\nonumber
\\
\end{eqnarray} 
Note that the indices in subscripts and those in parenthesis are equivalent, 
with the latter only distinguished by also labeling the nodes where data is 
distributed.  The size of the tensors is $N_t \times N_t \times N_t$, 
where $N_{t}$ is the number of momentum points times the number of discrete Matsubara
frequencies, i.e., $N_{t} = N_{k} \times N_{\omega}$. 
All indices are in modulo arithmetic at each of the $D+1$ dimensions, where $D=2$ (the cluster dimension), and the $``1"$ is for the Matsubara frequency. Because it takes many
iterations (up to a few hundred for low temperatures and strong coupling) 
to obtain converged solutions, the total number of tensor
rotations is significant and account for a large fraction of
the computational time.

We use the hybrid MPI/OpenMP model for the computations.
The rank three tensors are decomposed and evenly distributed into $N$
virtual nodes. Each virtual node consists of a few cores. The size of
a virtual node (i.e., the number of cores) is less than or equal to
the size of a physical node. Specifically, we slice the rank three
tensors to a set of two-dimensional arrays based on the index in 
parenthesis, e.g., $Q$ and $P-P^{\prime}$ for the left and right sides of
Eq.~\ref{rot1}, respectively. Then, each two dimensional matrix is assigned to a virtual node.  Since we have $N_t$
layers of two dimensional slices, the total number of virtual nodes also becomes
$N=N_t$. In this scenario, every rotation requires data communications
among all nodes. The following describes the data access patterns for our implementation of 
the tensor rotations.

\noindent \textbf{Step 1:} This step involves no MPI communication and
is done before any data is sent between nodes. The tensor elements are 
locally rearranged in order to collect specific elements to be grouped and
sent to designated destination nodes. The index in parenthesis of 
the tensors on the right of Eqs.~\ref{rot1}--\ref{rot4} represents the 
rank of a sending node in which a sliced two dimensional matrix resides. 
For rotations 1 and 2, rank of sending node $S$ is
\begin{equation}
S = P^{\prime}-P.
\label{rot12s}
\end{equation}
For rotations 3 and 4, $S$ is
\begin{equation}
S = P+P^{\prime}+Q.
\label{rot34s}
\end{equation}
Using Eqs.~\ref{rot12s} and \ref{rot34s}, and applying these to
the corresponding rotations, the two dimensional matrix elements are grouped based 
on the rank of destination node $Q$ from a given sending node $S$.
\begin{eqnarray}
{\rm rotation \: 1:}\ A_{P,Q} & = & \Phi(S)_{P,P+Q}  \label{step1a} \\
{\rm rotation \: 2:}\ A_{P,Q} & = & \Phi(S)_{-(P+S),P+Q}  \label{step1b} \\
{\rm rotation \: 3:}\ A_{P,Q} & = & \Phi(S)_{P+Q-S,-P}  \label{step1c} \\
{\rm rotation \: 4:}\ A_{P,Q} & = & \Phi(S)_{-(P+Q),-P}  \label{step1d}
\end{eqnarray}
Note that, here, $S$ is the node index (the index of the sender) and $P, Q \in
\{0,\ldots,N_t-1\}$, so the $P$ and $Q$ are the row and column indices of the 
matrix.  We assume column-major order data access in MPI data communications 
which distribute columns of matrix $\mathbf{A}$ to nodes of rank $Q$ in the 
next step.

\noindent \textbf{Step 2:} The columns of the two dimensional matrix $\mathbf{A}$
are distributed among all nodes.  At the sending nodes, each column of
$\mathbf{A}$ is sent to a destination node labeled by $Q$. The 
standard approach is to use \emph{MPI\_ALLTOALL}. However, as we show
later, this task can be done using different combinations of
point-to-point communications.\cite{mpi} In particular, non-blocking
communication protocols can be applied to overlap communications and
local computations. Overall, this procedure is applied to all the
tensor rotations and can be written as
\begin{equation}
B_{P,S} \textit{ at rank Q node: }  \leftarrow \: A_{P,Q} \textit{ at rank S node}.
 \label{step2}
\end{equation}
As shown in Eq.~\ref{step2}, the rank of destination nodes is
determined by the column index $Q$ of $\mathbf{A}$ in sending
nodes. The rank of sending nodes becomes column index $S$ of
$\mathbf{B}$ in the receiving nodes. The rank of sending nodes $S$
must be provided to receiving nodes in order to assign the correct
column index to the received messages.

\noindent \textbf{Step 3:} Once messages have arrived at the
destination nodes, the columns of the two dimensional matrix $\mathbf{B}$ are
rearranged to complete the tensor rotations.  
The column index of the rotated received matrix is related to the rank of the sending and 
receiving nodes by Eqs.~\ref{rot12s} and \ref{rot34s}.

Then, the
rotations are finalized by using the following relations
\begin{eqnarray}
{\rm rotation \: 1 , \: 2:} & \! & \Phi(Q)_{P,S+P} \longleftarrow B_{P,S} \label{step3ab} \\ 
{\rm rotation \: 3 , \: 4:} & \! & \Phi(Q)_{P,S-(P+Q)} \longleftarrow B_{P,S}, \label{step3cd}
\end{eqnarray}
where $Q$ is the index of a given receiving node and $P,S \in
\{0,\ldots,N_t-1\}$.  This step is a local process, i.e., no internode
communication is necessary.

\subsection*{Improving the Performance of Tensor Rotations}

While steps 1 and 3 are strictly local processes, step 2 is the 
only stage involving nonlocal MPI communications. The nature of the
collective communications among all nodes in step 2 makes it  
suited to the use of \emph{MPI\_ALLTOALL}. In such a case, step 2 can
start only after the completion of step 1. Because \emph{MPI\_ALLTOALL} 
is a blocking communication, step 3 must wait to start
until step 2 is finished. Therefore, the total elapsed time to
complete a tensor rotation is the sum of elapsed times of the three
steps. When the problem size is large, the communication efficiency of
\emph{MPI\_ALLTOALL} is reduced significantly due to the increased
network complexity associated with the bandwidth and latency among all
participating nodes. Our approach to handle these rotations more efficiently 
is to implement a latency hiding strategy by overlapping message
communications (step 2) and local computations (steps 1 and 3).

To enable this, we have developed our own version of a 
routine that performs communications from all nodes to all nodes. At a
basic level, the functionality of this routine is identical to that of
the generic \emph{MPI\_ALLTOALL} routine. However, our routine allows
further data manipulations such that local computations are embedded
between communications in the following way: On the sending node, the
first column of $\mathbf{A}$ is computed from the equations of step
1. Then, \emph{MPI\_ISEND} sends out the first column of $\mathbf{A}.$
While this column is being sent out, the next column of
$\mathbf{A}$ is prepared with step 1. This procedure is repeated until
all $N_t$ columns of $\mathbf{A}$, the group of the selected elements
from $\mathbf{\Phi}$, are sent out. This process overlaps steps 1 and
2. Latency hiding is also implemented in receiving nodes. We note that
the sending nodes are also receiving nodes. They only differ by
whether they are operating in the sending or the receiving {\it
mode}. On the receiving nodes, \emph{MPI\_IRECV} is set to receive
messages from arbitrary nodes by using \emph{MPI\_ANY\_SOURCE} as a
tag identifying the source of the message. For efficiency reasons,
\emph{MPI\_IRECV} is posted before \emph{MPI\_ISEND} of the sending
process. Then, \emph{MPI\_TEST} calls are used to check the completion
of the arrival of the message. Once message arrival is confirmed, the
rank of the node that sent this message can be identified by inquiring
using \emph{MPI\_STATUS}. This provides $S$ to assign to a
corresponding column and to be used in step 3. 
Since the message arrival is column-by-column, the processing of each column of
$\mathbf{B}$ continues to step 3 while the next column is
traveling through the network. This process is repeated until all columns
are completed.  This procedure completely overlaps steps 2 and 3.

Depending on the size of problem, it is desirable to define a virtual
node containing several cores (assuming multicore hardware
architecture) based on the memory availability per node. Among the cores,
MPI communications are assigned to one core. The other cores are
utilized by implementing OpenMP\cite{openmp} that parallelizes the
local computational tasks in a node to all cores within the
node. Thus, OpenMP thread depth is set to match with the total number
of cores per virtual node. Specifically, we applied the \emph{DO}
directive of OpenMP for iterations of index $P$ in the column
selection processes of steps 1 and 3.

\subsection*{Experimental Results}

We test the efficiency of this latency hiding scheme using a non-blocking protocol
against the standard \emph{MPI\_ALLTOALL}.  All the experimental
comparisons are conducted on the Cray XT5 (Jaguar) at the National
Center for Computational Sciences (NCCS) at the Oak Ridge National
Laboratory. Jaguar consists of 12 cores per node, with six cores per 
NUMA (Non-Uniform Memory Access) node, and two NUMAs per node. First, we discuss hardware-driven
constraints in implementing latency hiding. The non-blocking
\emph{MPI\_ISEND} does not check for the arrivals of messages. With
larger tensor size, the node usage and the size of individual columns
becomes large. The \emph{MPI\_ISEND} from all participating nodes
tries to dump a large column in each iteration. The next iteration
starts regardless of message arrivals in the receiving nodes. As a
consequence, a large amount of data rushes onto the network faster
than the data can be absorbed by the receiving nodes. Eventually, this
causes memory overflow to the system buffer assigned to the message
processing unit. To avoid this we have allocated more memory space to
the system buffer.

\begin{figure}[htc]
\includegraphics[scale=0.5]{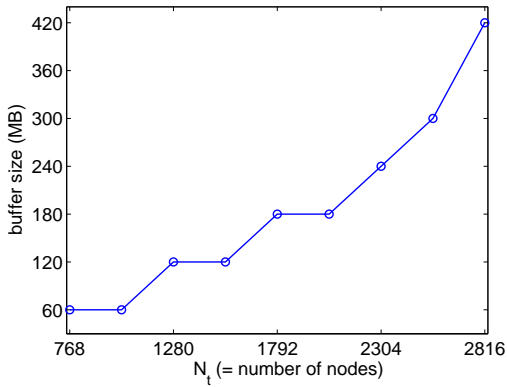}
\caption{(Color online) Required minimum buffer size to execute our
all-to-all routine; each node has 12 cores. \label{minbuffer}}
\end{figure} 

For simplicity, we assign one virtual node to a physical node. On the
Jaguar Cray XT5, this means one virtual node containing 12 cores. To
utilize all cores in a node, the value of OpenMP thread depth is set
to 12. We gradually increase the problem size $N_t$ until jobs end
with error indicating buffer overflow. Then, we set a higher buffer
size by controlling environmental variable
\emph{MPICH\_UNEX\_BUFFER\_SIZE}. For every incidence of error, we add
60 MB buffer size. The default value of
\emph{MPICH\_UNEX\_BUFFER\_SIZE} is 60 MB on JAGUAR XT5 (the total
number of cores is less than 50,000). The results are shown in the
Fig. ~\ref{minbuffer}.  Up to $N_t = 1024$, the 60 MB default buffer
size is enough to handle the data traffic. 
Increasing $N_t$ further forces us to use a larger buffer size. Overall, the
amount of added buffer size increases for larger problem sizes. We
note that the results presented in Fig.~\ref{minbuffer} are with the
maximum number of cores per a virtual node. Smaller core usage per
node alleviates the buffer restriction. For example, hardware setup
with a NUMA  node per virtual node consumes 
less buffer memory due to the reduced total number of physical nodes 
participating in internode communication. We did not observe buffer 
memory overflow with the generic blocking \emph{MPI\_ALLTOALL} routine.

\begin{figure}[htc]
\includegraphics[scale=0.55]{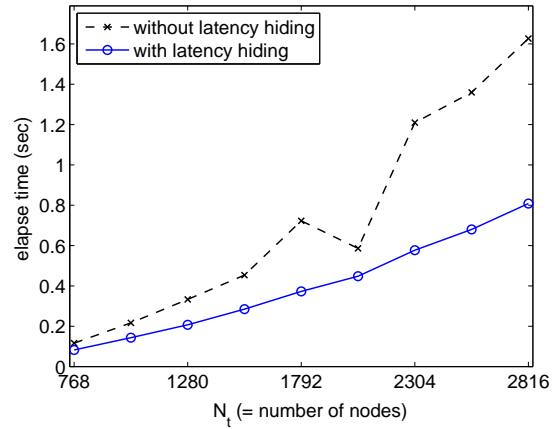}
\caption{(Color online) Time spent in data communication as a function of the number of computer nodes (12 processors per node). 
For large data sets each process sends messages to all the others, and  the communication time scales linearly 
with the number of processes. Latency hiding techniques that overlap the interprocessor communication with local 
computations yields a factor of two speedup when compared with the standard \emph{MPI\_ALLTOALL} implementations 
as the number of processors increases beyond 30,000.
\label{compare1}}
\end{figure}

The performance of the latency hiding approach is evaluated in terms of
wallclock time spent on a single tensor rotation and compared with the
case of the standard \emph{MPI\_ALLTOALL} applied for step 2. For
this, the elapsed time to complete the tensor rotation is averaged
over nine independent runs. Each run contains 40 repetitions of
identical tensor rotations. At the end of each run, the elapsed time
is also averaged for the 40 rotations. For all runs, we choose
rotation 1 and the minimum buffer sizes shown in the
Fig.~\ref{minbuffer} are assumed. The comparison results are shown
in Fig.~\ref{compare1}. Except for $N_t = 768$ and $2048$,
latency hiding outperforms the case without latency hiding in
significant amount. The performance differences are even higher for
$N_t \ge 2304$. For the \emph{MPI\_ALLTOALL} case, there is a sudden
speed-up at $N_t=2048$. We are exploring this behavior further.  We
believe that it is caused by changes in the data traffic 
controlled by the hardware.

Overall, latency hiding provides a higher speed-up for larger tensor
sizes and core count. From the general trends, it can
be expected that two-fold or more efficiency improvement for $N_t$
greater than $2816$ can be obtained by implementing latency hiding
with our non-blocking adaptation of the all-to-all routine for tensor
rotations.



\end{document}